\documentclass[aps,superscriptaddress,twocolumn,dvips]{revtex4}
\usepackage{feynmp}
\usepackage{amsmath}
\usepackage{amssymb}
\usepackage{epsfig}
\usepackage{graphicx}
\usepackage{subfigure}
\usepackage{hyperref}
\usepackage{bbm}
\usepackage{color}
\usepackage{longtable}
\usepackage{booktabs}
\usepackage{multirow}

\newcommand{\Rmnum}[1]{\expandafter\@slowromancap\romannumeral #1@}
\allowdisplaybreaks[3]
\begin{document}

\title{Topological quantum critical point in triple-Weyl semimetal:
non-Fermi-liquid behaviors and instabilities}

\author{Jing-Rong Wang}
\affiliation{Anhui Province Key Laboratory of Condensed Matter
Physics at Extreme Conditions, High Magnetic Field Laboratory of the
Chinese Academy of Sciences, Hefei, Anhui 230031, China}
\author{Guo-Zhu Liu}
\altaffiliation{Corresponding author: gzliu@ustc.edu.cn}
\affiliation{Department of Modern Physics, University of Science and
Technology of China, Hefei, Anhui 230026, China}
\author{Chang-Jin Zhang}
\altaffiliation{Corresponding author: zhangcj@hmfl.ac.cn}
\affiliation{Anhui Province Key Laboratory of Condensed Matter
Physics at Extreme Conditions, High Magnetic Field Laboratory of the
Chinese Academy of Sciences, Hefei, Anhui 230031, China}
\affiliation{Institute of Physical Science and Information
Technology, Anhui University, Hefei, Anhui 230601, China}

\begin{abstract}
We study the quantum critical phenomena emerging at the transition
from triple-Weyl semimetal to band insulator, which is a topological
phase transition described by the change of topological invariant.
The critical point realizes a new type of semimetal state in which
the fermion dispersion is cubic along two directions and quadratic
along the third. Our renormalization group analysis reveals that,
the Coulomb interaction is marginal at low energies and even
arbitrarily weak Coulomb interaction suffices to induce an infrared
fixed point. We compute a number of observable quantities, and show
that they all exhibit non-Fermi liquid behaviors at the fixed point.
When the interplay between the Coulomb and short-range four-fermion
interactions is considered, the system becomes unstable below a
finite energy scale. The system undergoes a first-order topological
transition when the fermion flavor $N$ is small, and enters into a
nematic phase if $N$ is large enough. Non-Fermi liquid behaviors are
hidden by the instability at low temperatures, but can still be
observed at higher temperatures. Experimental detection of the
predicted phenomena is discussed.
\end{abstract}

\maketitle


\section{Introduction}

Weakly interacting fermion systems are well described by Fermi
liquid (FL) theory \cite{GiulianiBook, ColemanBook}. Inter-particle
interactions induce a variety of continuous phase transitions, such
as magnetically ordered transition \cite{Sachdevbook, Lohneysen07},
charge-density-wave transition \cite{Fradkin15}, and nematic
transition \cite{Fradkin10}. This type of transition happens between
two phases with distinct symmetries. According to
Ginzburg-Landau-Wilson (GLW) theory \cite{Sachdevbook}, one can
define a local order parameter to describe such a transition. At the
quantum critical point (QCP), fermionic excitations interact
strongly with the quantum fluctuation of order parameter, which
often leads to breakdown of FL theory \cite{Sachdevbook,
Lohneysen07, Fradkin15, Fradkin10, Varma02}. However, not all phase
transitions are classified by the change of symmetries. Transitions
may occur between phases that respect the same symmetries but are
topologically distinct \cite{WenXG04, WenXG17}. These transitions
are beyond the paradigm of GLW theory, and could be described by the
change of global topological invariant.

In recent years, a plethora of semimetal (SM) materials have been
discovered and extensively investigated \cite{Kotov12, Vafek14,
Wehling14, Wan11, Weng16, FangChen16, Yan17, Hasan17, Armitage18,
SonBurkov, BurkovNandy}. Under suitable conditions, SM states emerge
at the toplogical QCP (TQCP) between two topologically distinct
phases \cite{Murakami07, Goswami11, Montambaux09, Isobe16,
RoyFoster18, Cho16, WangLiuZhang17A, AnhYang17, Yang13, Yang14A,
Yang14B, Mirakami17, Roy18B}. For instance, three-dimensional (3D)
topological insulator (TI) can be turned into a band insulator (BI),
with gapless 3D Dirac SM (DSM) state emerging at TI-BI QCP
\cite{Murakami07, Goswami11, Xu11, Sato11}.

Inter-particle interaction effect at TQCP is a nontrivial issue and
has attracted particular interest recently \cite{Goswami11, Isobe16,
Cho16, WangLiuZhang17A, Yang14B}. The fermions do not couple to any
order parameter at the TQCP. Their properties are determined mainly
by the Coulomb interaction, which is long-ranged because the density
of states (DOS) vanishes at Fermi level. The effects of Coulomb
interaction are diverse in various SM systems, depending sensitively
on the spatial dimensionality and the fermion energy dispersion
\cite{Goswami11, Isobe16, Cho16, WangLiuZhang17A, Yang14B, Moon13,
Herbut14, Janssen17, Lai15, Jian15, Huh16, WangLiuZhang17B, Zhang17,
WangLiuZhang18}.

\begin{figure}[htbp]
\center
\includegraphics[width=3.33in]{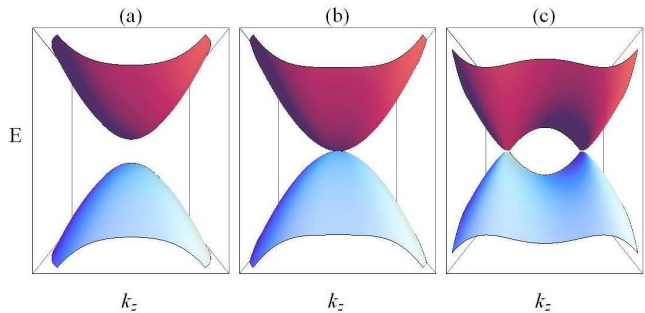}
\caption{As a triple-WSM is converted into a normal BI, two
band-touching points merge to one single point at the TQCP. The
fermion dispersion can be universally expressed in the form $E = \pm
\sqrt{B^{2}k_{\bot}^{6} + \left(Ak_{z}^{2}-\Delta\right)^{2}}$. (a)
BI with $\Delta < 0$; (b) TQCP with $\Delta=0$; (c) Triple-WSM with
$\Delta > 0$. \label{Fig:Merge}}
\end{figure}

In this article, we consider the transition between topological
triple-Weyl SM (WSM) and BI. In a triple-WSM, the fermion energy
spectrum disperses cubically along two directions and linearly along
the third one, and the monopole charges of one pair of Weyl points
are $\pm 3$ \cite{WangLiuZhang17B, Zhang17, WangLiuZhang18,
XuFang11, Fang12, Ahn17, Park17, Roy17, Huang17, Hayata17, Gorbar17,
Ezawa17, Dantas18}. In the process of turning triple-WSM into BI,
two Weyl points carrying opposite monopole charges merge to form one
single band-touching point that carries zero monopole charge. This
is a topological quantum phase transition (TQPT) since the monopole
charge is finite in one phase and zero in the other. The merging
process is shown in Fig.~\ref{Fig:Merge}. At the TQCP, the fermion
dispersion is cubic along two directions and quadratic along the
third, which defines a new gapless SM state.

We investigate the influence of long-range Coulomb interaction on
the triple-WSM-BI TQCP by carrying out an extensive renormalization
group (RG) analysis. The Coulomb interaction is found to be marginal
in the low-energy regime. The system flows to an infrared fixed
point, at which the interaction strength takes a finite value. We
compute a number of observable quantities and demonstrate that they
all exhibit striking non-FL (NFL) behaviors at such a fixed point.

\begin{figure}[htbp]
\center
\includegraphics[width=2.9in]{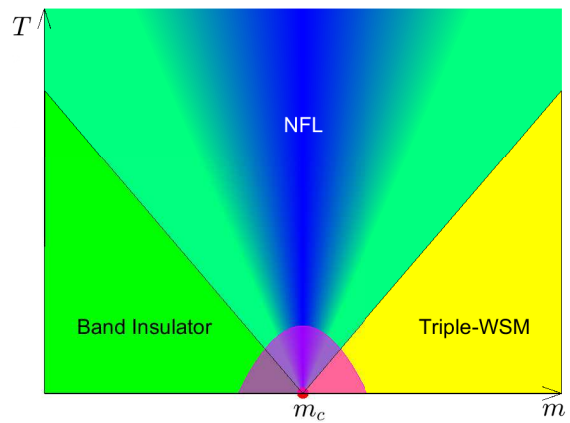}
\caption{Phase diagram on $m$-$T$ plane, where $m$ tunes the
transition and $T$ is temperature. The TQCP ($m_c$) is broadened
into a NFL-like quantum critical regime at finite $T$. Magenta dome
appearing at low $T$ represents either the formation of nematic
order or the occurrence of first-order TQPT.
\label{Fig:Phasediagram}}
\end{figure}

We then study the possible instabilities of this fixed point by
analyzing the interplay between the Coulomb interaction and some
sorts of short-range four-fermion interactions. Our finding is that,
the system undergoes a first order TQPT when the fermion flavor $N$
is small and develops a nematic order if $N$ is sufficiently large.
The NFL behaviors are hidden at low temperatures due to the
instability, but still have observable effects at higher
temperatures. All these results, summarized in the schematic phase
diagram given by Fig.~\ref{Fig:Phasediagram}, can be verified by
experiments.

The rest of the paper will be organized as follows. In
Sec.~\ref{Sec:Model}, the effective low-energy action of the TQCP is
introduced and described. In Sec.~\ref{Sec:RGAnalysis}, the coupled
RG equations for model parameters are presented, and the RG
solutions are analyzed. The NFL behaviors of observable quantities
are discussed in Sec.~\ref{Sec:ObsQuan}. The possible
phase-transition instabilities induced by the Coulomb and
short-range interactions are investigated in
Sec.~\ref{Sec:InstabiltiesCoulomb}. In Sec.~\ref{Sec:SummaryDiss},
the experimental detection of our theoretical predictions is briefly
discussed. All of calculational details are presented in the
Appendices.

\section{The model \label{Sec:Model}}

At the triple-WSM-BI TQCP, the free Hamiltonian of
the gapless fermions takes the form
\begin{equation}
H_{f} = \sum_{i=1}^{3}\int
d^3\mathbf{x} \psi_{a}^{\dag}(\mathbf{x})
d_{i}(\mathbf{x})\sigma_{i} \psi_{a}(\mathbf{x}),
\end{equation}
where $d_{1}(\mathbf{x}) = Bi(\partial_{x}^{3} -
3\partial_{x}\partial_{y}^{2})$, $d_{2}(\mathbf{x}) =
Bi(\partial_{y}^{3} -
3\partial_{y}\partial_{x}^{2})$, and $d_{3}(\mathbf{x}) =
-A\partial_{z}^{2}$. The field operator $\psi_{a}$ is a
two-component spinor, and repeated indices of $a$ implies the
summation from $1$ to $N$, where $N$ is the fermion flavor.
$\sigma_{1,2,3}$ are the Pauli matrices, and $B$ and $A$ are model
parameters. The fermion energy dispersion is
$E=\pm\sqrt{B^2k_{\bot}^{6}+A^{2}k_{z}^{4}}$,
where $k_{\bot}^{2}=k_{x}^{2}+k_{y}^{2}$. The Coulomb interaction can be
described by
\begin{eqnarray}
H_{C} = \frac{1}{4\pi}\int
d^3\mathbf{x}d^3\mathbf{x}'\rho_{a}(\mathbf{x})\frac{e^{2}}{\epsilon
\left|\mathbf{x}-\mathbf{x}'\right|}\rho_{a}(\mathbf{x}'),
\end{eqnarray}
where $\rho_{a}(\mathbf{x})=\psi_{a}^{\dag}(\mathbf{x})\psi_{a}(\mathbf{x})$
is fermion density operator, $e$ electric charge, and $\epsilon$
dielectric constant. One can treat Coulomb interaction by defining
an auxiliary boson field $\phi$ via the Hubbard-Stratonovich
transformation \cite{Goswami11, Isobe16, Cho16, Yang14B,
WangLiuZhang17B, Zhang17, Moon13, Herbut14, Janssen17, Lai15,
Jian15, Huh16}.

The total effective action is formally expressed as
$S = S_{\psi}+S_{\phi} + S_{\psi\phi}$, where
\begin{eqnarray}
S_{\psi} &=& \int\frac{d\omega}{2\pi}\frac{d^3\mathbf{k}}{(2\pi)^{3}}
\psi_{a}^{\dag}\left(i\omega-\mathcal{H}_{f}(\mathbf{k})\right)\psi_{a} ,
 \\
S_{\phi} &=& \int\frac{d\omega}{2\pi}\frac{d^3\mathbf{k}}{(2\pi)^{3}}
\phi\left(k_{\bot}^{2}+\eta k_{z}^{2}\right)\phi,
\\
S_{\psi\phi} &=& ig\int d\tau d^3\mathbf{x} \phi\psi_{a}^{\dag}
\psi_{a}.
\end{eqnarray}
Here, $g = \sqrt{4\pi}e/\sqrt{\epsilon}$. Because the fermion
dispersion is anisotropic, the three momentum components need to be
re-scaled differently. To facilitate RG calculations, we have
introduced a parameter $\eta$. The bare fermion propagator is given
by
\begin{eqnarray}
G_{0}(i\omega,\mathbf{k}) =\frac{1}{ i\omega -
d_{1}(\mathbf{k})\sigma_{1} - d_{2}(\mathbf{k})\sigma_{2} -
d_{3}(\mathbf{k})\sigma_{3}},
\end{eqnarray}
where $d_{1}(\mathbf{k}) = B\left(k_{x}^{3}-3k_{x}k_{y}^{2}\right)$,
$d_{2}(\mathbf{k}) = B\left(k_{y}^{3}-3k_{y}k_{x}^{2}\right)$, and
$d_{3}(\mathbf{k}) = Ak_{z}^{2}$. The bare boson propagator reads
\begin{eqnarray}
D_{0}(\mathbf{q}) =
\frac{1}{q_{\bot}^{2}+\eta q_{z}^{2}},
\end{eqnarray}
which is clearly long-ranged.

\section{Renormalization group analysis \label{Sec:RGAnalysis}}

We now employ the RG method \cite{Shankar94} to study the Coulomb
interaction on the low-energy properties of fermions. The momentum
shell to be integrated out is chosen as $b\Lambda <
\sqrt{B^{2}k_{\bot}^{6}+A^{2}k_{z}^{4}} < \Lambda$, where $\Lambda$
is some high-energy cutoff and $b = e^{-\ell}$ with $\ell$ being the
flow parameter. After performing tedious calculations, we obtain the
fermion self-energy
\begin{eqnarray}
\Sigma(i\omega,\mathbf{k}) \approx
-\left[d_{1}(\mathbf{k})\sigma_{1} +
d_{2}(\mathbf{k})\sigma_{2}\right]C_{2}\ell - Ak_{z}^{2}
\sigma_{3}C_{3}\ell. \label{Eq:FermionSelfEnergyMainText}
\end{eqnarray}
The expressions of $C_{2}$ and $C_{3}$ are complicated and can be
found in Appendix~\ref{App:FermionSelfEnergy}. According to
Appendix~\ref{App:BosonSelfEnergy}, the boson self-energy calculated
in the static limit is given by
\begin{eqnarray}
\Pi(\mathbf{q}) \approx q_{\bot}^{2}C_{\bot}\ell +
q_{z}^{2}C_{z}\ell, \label{Eq:BonsonSelfEnergyMainText}
\end{eqnarray}
where
\begin{eqnarray}
C_{\bot} &=& \frac{9g^{2}}{40\pi^{2} \sqrt{A\Lambda}},
\\
C_{z}&=& \frac{\Gamma\left(\frac{3}{4}\right)\Gamma
\left(\frac{4}{3}\right)g^{2}\sqrt{A}}{24\pi^{2}\Gamma
\left(\frac{25}{12}\right)B^{2/3}\Lambda^{5/6}}.
\end{eqnarray}
According to the calculations presented in
Appendix~\ref{App:DerivationRGQuation}, the coupled RG equations for
model parameters are
\begin{eqnarray}
\frac{dB}{d\ell}&=&C_{2}B, \label{Eq:VRGB}
\\
\frac{dA}{d\ell}&=&C_{3}A, \label{Eq:VRGA}
\\
\frac{d\alpha}{d\ell} &=& \left(\frac{1}{2} -
NC_{\bot}-\frac{C_{3}}{2}\right)\alpha, \label{Eq:VRGAlpha}
\\
\frac{d\beta}{d\ell} &=& \left(\frac{5}{6} +\frac{1}{2}C_{3} -
\frac{2}{3}C_{2}-N\beta\right)\beta, \label{Eq:VRGBeta}
\\
\frac{d\eta}{d\ell} &=& \left(-\frac{1}{3} -
NC_{\bot}+N\beta\right)\eta, \label{Eq:VRGEta}
\\
\frac{d\zeta}{d\ell} &=& \left(-\frac{1}{3}+\frac{2}{3}C_{2} -
C_{3}-NC_{\bot}+N\beta\right)\zeta. \label{Eq:VRGZeta}
\end{eqnarray}
Here, we define three new parameters: $\alpha = 5C_{\bot}/9\pi$,
$\beta = C_{z}/\eta$, and $\zeta = \eta B^{2/3}\Lambda^{1/3}/A$. The
parameter $\alpha$ represents the effective Coulomb interaction
strength, whereas the parameter $\beta$ is related to the dynamical
screening of Coulomb interaction. In our RG calculations, $1/N$
serves as a formal control parameter.

\begin{figure}[htbp]
\center
\includegraphics[width=3.36in]{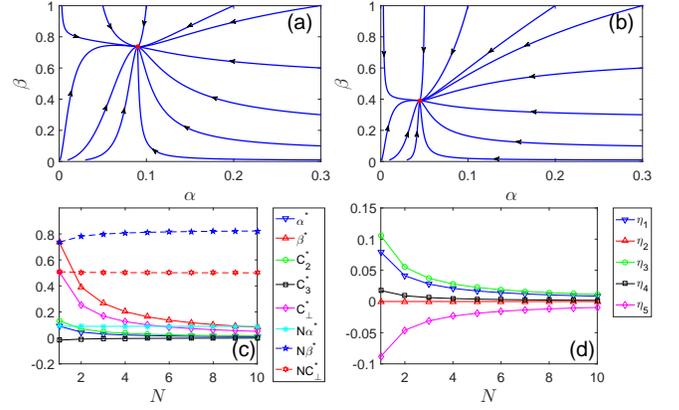}
\caption{Flow diagram on $\alpha$-$\beta$ plane for $N=1$ is given
in (a) and for $N=2$ in (b). Red point $(\alpha^{*},\beta^{*})$ is
an infrared fixed point. (c) and (d) show the $N$-dependence of some
parameters. \label{Fig:FlowDiagramNoFF}}
\end{figure}

We plot the flow diagrams on the $\alpha$-$\beta$ plane in
Figs.~\ref{Fig:FlowDiagramNoFF}(a) and \ref{Fig:FlowDiagramNoFF}(b)
based on the solutions of Eqs.~(\ref{Eq:VRGB})-(\ref{Eq:VRGZeta}).
There is a stable infrared fixed point $(\alpha^{*},\beta^{*})$. In
the lowest energy limit, $C_{2}$, $C_{3}$, and $C_{\bot}$ approach
to three constants $C_{2}^{*}$, $C_{3}^{*}$, and $C_{\bot}^{*}$,
respectively. The values of the constants $\alpha^{*}$, $\beta^{*}$,
$C_{2}^{*}$, $C_{3}^{*}$, and $C_{\bot}^{*}$ are shown in
Fig.~\ref{Fig:FlowDiagramNoFF}(c), and can be found in
Appendix~\ref{App:DerivationRGQuation}. The constant $C_{3}^{*} <
0$, but other constants are positive. At low energies, the
parameters $B$ and $A$ behave as
\begin{eqnarray}
B&\sim& e^{C_{2}^{*}\ell},
\\
A&\sim& e^{C_{3}^{*}\ell}.
\end{eqnarray}
As the energy decreases, $B$ increases
indefinitely, while $A$ goes to zero quickly. Since $B$ and $A$
enter into almost all the observable quantities, such a
renormalization will certainly have observable effects.

Including polarization $\Pi(\mathbf{q})$ into the boson propagator
leads to renormalized Coulomb interaction
\begin{eqnarray}
D(\mathbf{q}) = \frac{1}{D_{0}^{-1}(\mathbf{q})+\Pi(\mathbf{q})}
\sim \frac{1}{q_{\bot}^{2-3NC_{\bot}^{*}} + q_{z}^{2-2N\beta^{*}}}.
\end{eqnarray}
Since $C_{\bot}^{*}$ and $\beta^{*}$ are both positive, the Coulomb
interaction is substantially screened.

\section{Observable Quantities \label{Sec:ObsQuan}}

The infrared fixed point $(\alpha^{*},\beta^{*})$ is characterized
by the emergence of striking NFL behaviors. To demonstrate this, we
will compute a number of observable quantities, including the
fermion DOS, specific heat, compressibility, dynamical
conductivities, and diamagnetic susceptibilities. The detailed
calculations of observable quantities can be found in
Appendices~\ref{App:ObserQuanFree} and \ref{App:ObserQuanInteract}.

\begin{table*}[htbp]
\caption{The energy or temperature dependence of observable
quantities at TQCP between triple-WSM and BI. The
interaction-induced extra exponents are shown in
Fig.~\ref{Fig:FlowDiagramNoFF}(d) and also in
Appendix~\ref{App:ObserQuanInteract}}.
\label{Table:ObservableQuantities} \vspace{-0.3cm}
\begin{center}
\setlength{\tabcolsep}{5.2mm}{
\begin{tabular}{|c|c|c|c|c|c|c|c|}
\hline\hline  & $\rho(\omega)$ & $C_{v}(T)$ & $\kappa(T)$ &
$\sigma_{\bot\bot}(\omega)$ & $\sigma_{zz}(\omega)$ &
$\chi_{D}^{\bot}(T)$ & $\chi_{D}^{z}(T)$
\\
\hline Free & $\omega^{1/6}$ & $T^{7/6}$ & $T^{1/6}$ &
$\omega^{1/2}$ & $\omega^{1/6}$ & $T^{1/2}$ & $T^{5/6}$
\\
\hline Interacting & $\omega^{1/6+\eta_{1}}$ & $T^{7/6+\eta_{1}}$ &
$T^{1/6+\eta_{1}}$ & $\omega^{1/2+\eta_{2}}$ &
$\omega^{1/6+\eta_{3}}$ & $T^{1/2+\eta_{4}}$ & $T^{5/6+\eta_{5}}$
\\
\hline\hline
\end{tabular}}
\end{center}
\end{table*}

We first consider the non-interacting limit. It is easy to get the
fermion DOS
\begin{eqnarray}
\rho(\omega) = \frac{c_{1}}{\sqrt{A}B^{2/3}}\omega^{1/6}.
\label{Eq:DOSFreeMainText}
\end{eqnarray}
Apparently, $\rho(\omega)$ vanishes in the limit $\omega\rightarrow
0$, which is a common feature of most SMs and renders that the
Coulomb interaction is long-ranged. The specific heat $C_{v}(T)$ and
compressibility $\kappa(T)$ are
\begin{eqnarray}
C_{v}(T) &=& \frac{c_{2}}{\sqrt{A}B^{2/3}}T^{7/6}, \label{Eq:CvFreeMainText}
\\
\kappa(T) &=& \frac{c_{3}}{\sqrt{A}B^{2/3}}T^{1/6},
\label{Eq:CompressibilityFreeMainText}
\end{eqnarray}
respectively. The dynamical conductivity within $x$-$y$ plane,
$\sigma_{\bot\bot}(\omega)$, and the one along $z$-axis,
$\sigma_{zz}(\omega)$, are
\begin{eqnarray}
\sigma_{\bot\bot}(\omega) &=& \frac{c_{4}e^{2}}{
\sqrt{A}}\omega^{1/2}, \label{Eq:SigmaBotFreeMainText}
\\
\sigma_{zz}(\omega) &=& \frac{c_{5}\sqrt{A}e^{2}}{
B^{2/3}}\omega^{1/6}. \label{Eq:SigmazzFreeMainText}
\end{eqnarray}
For the external fields applied within the $x$-$y$ plane and along
the $z$-axis, the diamagnetic susceptibilities are respectively
given by
\begin{eqnarray}
\Delta\chi_{D}^{\bot}(T) &=& c_{6}\sqrt{A}e^{2}T^{1/2},
\label{Eq:SuspBotFreeMainText}
\\
\Delta\chi_{D}^{z}(T) &=& c_{7}\frac{B^{2/3}
e^{2}}{\sqrt{A}}T^{5/6}, \label{Eq:SuspzzFreeMainText}
\end{eqnarray}
where $\Delta\chi_{D}^{\bot}(T) = \chi_{D}^{\bot}(T) -
\chi_{D0}^{\bot}$ and $\Delta\chi_{D}^{z}(T) =
\chi_{D}^{z}(T)-\chi_{D0}^{z}$, with $\chi_{D0}^{\bot}$ and
$\chi_{D0}^{z}$ being the residual values of the $T\rightarrow 0$
limit. The values of constants $c_{1}$, $c_{2}$, ..., $c_{7}$ are
given in Appendix~\ref{App:ObserQuanFree}.

We then incorporate the corrections induced by the Coulomb
interaction. Making use of the RG results of $B(\ell)$ and
$A(\ell)$, and then employing the scaling relation $\omega =
\omega_{0}e^{-\ell}$ or $T=T_{0}e^{-\ell}$, we can obtain the
renormalized observable quantities. First of all, the DOS becomes
\begin{eqnarray}
\rho(\omega)\sim\omega^{1/6+\eta_{1}}, \label{Eq:DOSIntMainText}
\end{eqnarray}
where $\eta_{1} > 0$. Comparing Eq.~(\ref{Eq:DOSFreeMainText}) to
Eq.~(\ref{Eq:DOSIntMainText}), we find that Coulomb interaction
effectively suppresses fermion DOS. Interaction corrections modify
specific heat and compressibility to
\begin{eqnarray}
C_{v}(T)&\sim& T^{7/6+\eta_{1}}, \label{Eq:CvIntMainText}
\\
\kappa(T)&\sim& T^{1/6+\eta_{1}}. \label{Eq:CompressibilityIntMainText}
\end{eqnarray}
These two quantities are also suppressed. The extra exponent
$\eta_1$ appearing in $\rho(\omega)$, $C_{v}(T)$, and $\kappa(T)$ is
exactly the same, which reflects the fact that $\rho(\omega)$,
$C_{v}(T)$, and $\kappa(T)$ display the same dependence on $B$ and
$A$, as shown in Eqs.~(\ref{Eq:DOSFreeMainText}),
(\ref{Eq:CvFreeMainText}), and (\ref{Eq:CompressibilityFreeMainText}).

According to Eqs.~(\ref{Eq:SigmaBotFreeMainText})-
(\ref{Eq:SuspzzFreeMainText}), the charge $e$ enters into the
expressions of dynamical conductivities and diamagnetic
susceptibilities. The RG flow of $e$ naturally affects the
low-energy properties of these quantities. From the
$\ell$-dependence of $B$, $A$, and $e$, we find that
\begin{eqnarray}
\sigma_{\bot\bot}(\omega) &\sim& \omega^{1/2+\eta_{2}},
\label{Eq:SigmaBotIntMainText}
\\
\sigma_{zz}(\omega)&\sim& \omega^{1/6+\eta_{3}},
\label{Eq:SigmazzIntMainText}
\end{eqnarray}
where $\eta_{2}=0$ and $\eta_{3}$ is a positive constant. Notice
that the $\omega$-dependence of $\sigma_{\bot\bot}$ is unchanged.
The reason is that, $e^{2}$ and $\sqrt{A}$ are renormalized in
precisely the same way by Coulomb interaction, which ensures that
$\sigma_{\bot \bot}\propto e^{2}/\sqrt{A}$ remains intact.
Eq.~(\ref{Eq:SigmazzFreeMainText}) and Eq.~(\ref{Eq:SigmazzIntMainText})
indicate that $\sigma_{zz}$ is suppressed by Coulomb interaction.

After including the interaction corrections, the diamagnetic
susceptibilities become
\begin{eqnarray}
\Delta\chi_{D}^{\bot}(T) &\sim& T^{1/2+\eta_{4}},
\\
\Delta\chi_{D}^{z}(T)&\sim& T^{5/6+\eta_{5}},
\end{eqnarray}
where $\eta_{4} > 0$ but $\eta_{5} < 0$.

We summarize the results for observable quantities in
Table~\ref{Table:ObservableQuantities}. The power-law corrections
indicate that Coulomb interaction induces typical NFL behaviors.

\section{Instability driven by Coulomb interaction \label{Sec:InstabiltiesCoulomb}}

An interesting consequence of the Coulomb interaction is to
dynamically generate some types of short-range four-fermion coupling
\cite{Herbut14, Janssen17, Juricic09, Roy16}. If the four-fermion
interaction flows to strong coupling regime, the system would become
unstable. Motivated by previous works on Luttinger SM
\cite{Herbut14, Janssen17}, we now study the possible instabilities
driven by Coulomb interaction through analyzing the interplay of
Coulomb and four-fermion interactions at the TQCP under
consideration.

There are four types of four-fermion interactions:
$\lambda_{1}\left(\psi^{\dag}\sigma_{1}\psi\right)^{2}$,
$\lambda_{2}\left(\psi^{\dag}\sigma_{2}\psi\right)^{2}$,
$\lambda_{3}\left(\psi^{\dag}\sigma_{3}\psi\right)^{2}$, and
$\lambda_{0}\left(\psi^{\dag}\psi\right)^{2}$. After decoupling each
four-fermion coupling term, one obtains a fermion bilinear. Each
fermion bilinear has its own meaning. Below, we study these four
cases separately in order.

\begin{figure}[htbp]
\center
\includegraphics[width=3.36in]{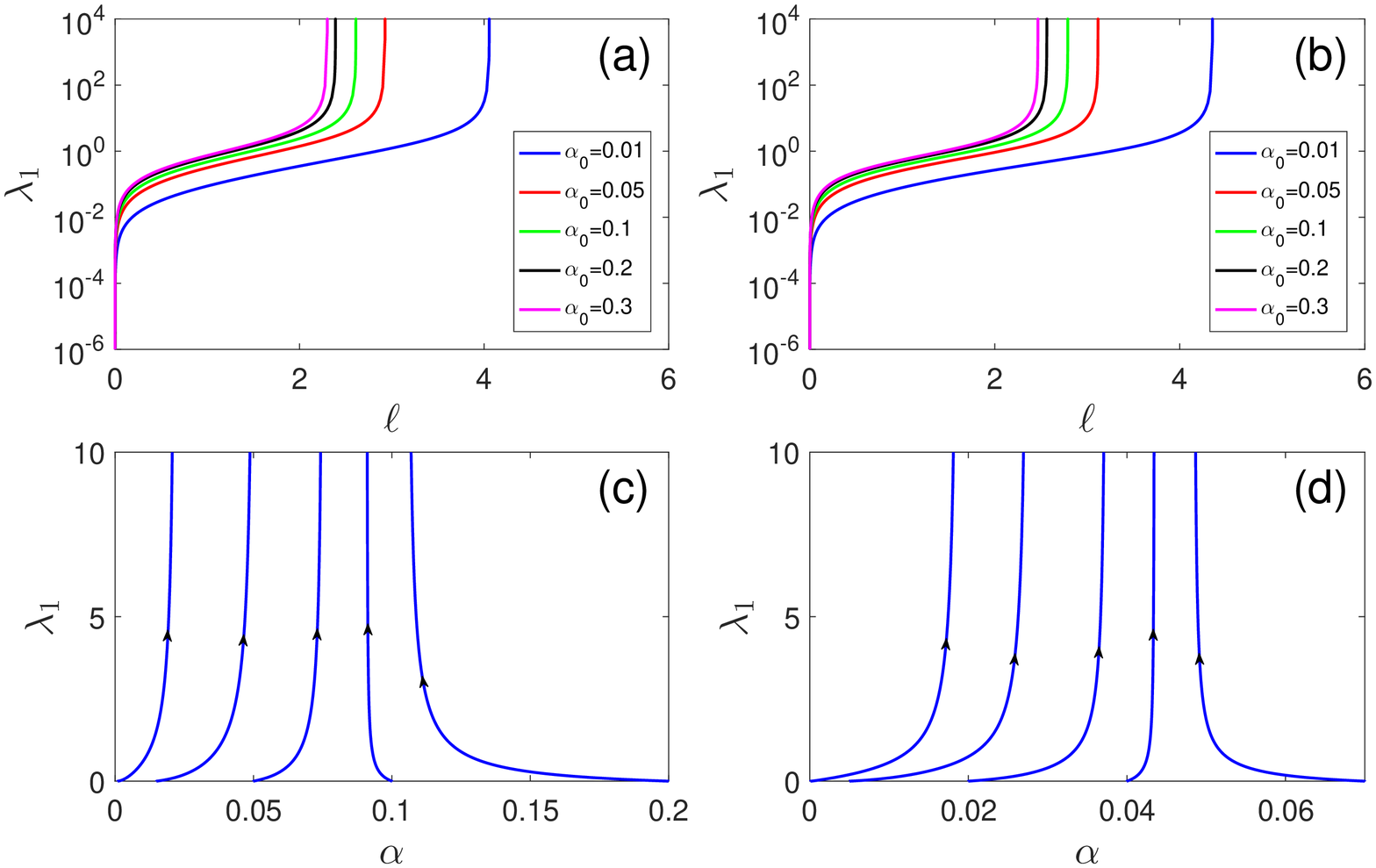}
\caption{Dependence of $\lambda_{1}$ on $\ell$ for $N=1$ is shown in
(a) and for $N=2$ in (b). Flow diagram on $\alpha$-$\lambda_{1}$
plane for $N=1$ is shown in (c) and for $N=2$ in (d). Here,
$\beta_{0}=0.1$. \label{Fig:FlowDiagramFF}}
\end{figure}

We first consider the following coupling
\begin{eqnarray}
S_{\psi^{4}} = \frac{1}{N}\int d\tau d^3\mathbf{x} \lambda_{1}
\left(\psi_{a}^{\dag}\sigma_{1}\psi_{a}\right)^{2}.
\end{eqnarray}
According to the RG analysis shown in
Appendix~\ref{App:InterplayCoulombFourFermion}, its coupling
parameter $\lambda_1$ satisfies the flow equation
\begin{eqnarray}
\frac{d\lambda_{1}}{d\ell} = -\frac{1}{6}\lambda_{1} + \left(1 -
\frac{2}{5N}\right)\lambda_{1}^{2} +F_{A}\lambda_{1}+F_{B},
\end{eqnarray}
where the redefinition
$\frac{5\Gamma\left(5/4\right)\Gamma\left(4/3\right)
\Lambda^{1/6}}{12 \pi^{2}\Gamma\left(19/12\right)
\sqrt{A}B^{2/3}}\lambda_{1} \rightarrow\lambda_{1}$ has been
employed. The factors $F_{A}$ and $F_{B}$ are induced by Coulomb
interaction. The flows of $\lambda_1$ are presented in
Fig.~\ref{Fig:FlowDiagramFF}. We observe from
Figs.~\ref{Fig:FlowDiagramFF}(a) and \ref{Fig:FlowDiagramFF}(b)
that, $\lambda_{1}(\ell)$ develops a finite value from zero as
$\ell$ increases and $\lambda_{1}(\ell)$ diverges as $\ell
\rightarrow \ell_{c}$ with $\ell_c$ being finite. The flow diagrams
plotted in Figs.~\ref{Fig:FlowDiagramFF}(c) and
\ref{Fig:FlowDiagramFF}(d) show that this conclusion holds even when
the initial value of $\alpha$ is arbitrarily small. As a result of
the runaway flow of $(\psi^{\dag} \sigma_{1}\psi)^{2}$, the
corresponding fermion bilinear acquires a finite expectation value,
namely $\Delta_{1} \equiv \langle \psi^{\dag}\sigma_{1}\psi \rangle
\neq 0$. The original band-touching point is split into three
different points: $\left(-\left(\Delta_{1}/B\right)^{1/3}, 0,
0\right)$ and $\left(\frac{1}{2}\left(\Delta_{1}/B\right)^{1/3},
\pm\frac{\sqrt{3}}{2}\left(\Delta_{1}/B\right)^{1/3}, 0\right)$. One
can identify $\Delta_{1}$ as a nematic order parameter, because the
positions of these points break the equivalence between $x$- and
$y$-axis. Thus, the system enters into a nematic state. Around these
three points, the fermion dispersion can be expressed as
\begin{eqnarray}
E=\pm\sqrt{9B^{2/3}\Delta_{1}^{4/3}K_{\bot}^{2}+A^{2}K_{z}^{4}},
\end{eqnarray}
which is linear along two directions and quadratic along the third.
$\mathbf{K}$ is  the momentum relative to the new touching points.
Such a dispersion is the same as that of an anisotropic-WSM
\cite{Yang13,Yang14B}, in which the Coulomb interaction is
irrelevant at low energies, and the DOS, specific heat, and
compressibility behave as $\rho(\omega)\sim\omega^{3/2}$,
$C_{v}(T)\sim T^{5/2}$, and $\kappa(T)\sim T^{3/2}$, respectively.
The mean value $\Delta_{1}$ is nonzero only at low $T$, and is
destroyed as $T$ grows. Therefore, the NFL behaviors caused by
Coulomb interaction are hidden at low $T$ by the nematic order, but
re-appear at higher $T$ once the nematic order is destroyed by
thermal fluctuation. The critical temperature $T_{n}$ can be roughly
estimated by $\Lambda e^{-\ell_{c}}$.

Similar to $\langle \psi^{\dag}\sigma_{1}\psi \rangle$, a nonzero
$\Delta_2 \equiv \langle \psi^{\dag}\sigma_{2}\psi \rangle$ also
splits the original band-touching point into three new points, which
are located at $(0,-\left(\Delta_{2}/B\right)^{1/3}, 0)$ and
$(\pm\frac{\sqrt{3}}{2}\left(\Delta_{2}/B\right)^{1/3},
\frac{1}{2}\left(\Delta_{2}/B\right)^{1/3}, 0)$. Thus, $\langle
\psi^{\dag}\sigma_{2}\psi \rangle$ also corresponds to nematic order
parameter.

If $\Delta_3 \equiv \langle \psi^{\dag}\sigma_{3}\psi \rangle \neq
0$, the parameter $\Delta$ (defined in Fig.~\ref{Fig:Merge}) becomes
finite, i.e., $\Delta = -\Delta_{3}$, at the TQCP. Consequently, the
transition between triple-WSM and BI becomes first-order
\cite{Roy17}, and the original TQCP is eliminated at low $T$. At
higher $T$, thermal fluctuation forces $\Delta_{3}$ to vanish, and
the effective fermion dispersion is still
$E=\pm\sqrt{B^2k_{\bot}^{6}+A^{2}k_{z}^{4}}$. The Coulomb
interaction leads to NFL behaviors at high $T$, which can be
explored by measuring the observable quantities of
Table~\ref{Table:ObservableQuantities}.

If $\Delta_{0} \equiv \langle \psi^{\dag}\psi \rangle \neq 0$, the
Fermi surface moves away from the band-touching point to a new
level, since $\langle \psi^{\dag}\psi \rangle$ plays the role of a
finite chemical potential.

As demonstrated in Appendix~\ref{App:InterplayCoulombFourFermion},
$\lambda_{2}$ and $\lambda_{3}$ are both generated from zero, and
both diverge at some finite energy scale. However, $\lambda_{0}$
grows from zero and finally approaches to a constant in the lowest
energy limit, which means that $\Delta_{0} \equiv 0$. Therefore,
there are only two different instabilities, namely the formation of
nematicity and the occurrence of first-order TQPT. The instability
that would shift the Fermi level is excluded.

A natural question arises as to which out of the two instabilities
is more favorable. As shown in
Appendix~\ref{App:InterplayCoulombFourFermion}, our numerical
studies reveal that, within a wide range of the initial value of
$\alpha$,  $\lambda_{3}$ diverges most rapidly for $N=1,2$, whereas
$\lambda_{1}$ and $\lambda_{2}$ diverge more quickly than
$\lambda_{3}$ if $N \geq 3$. We thus conclude that, first-order TQPT
occurs for $N=1,2$, but gapless nematic state is realized for $N
\geq 3$.

\section{Summary and discussion \label{Sec:SummaryDiss}}

In summary, we have investigated the quantum critical phenomena of
the TQCP between the topological triple-WSM and the normal BI, and
found that the long-range Coulomb interaction leads to a NFL
infrared fixed point. This infrared fixed point can be
experimentally probed by measuring the energy or temperature
dependence of a number of observable quantities. We also have
considered the interplay between Coulomb interaction and short-range
interactions. Such an interplay can either causes a long-range
nematic order or drive a first-order transition, depending on the
value of the fermion flavor at TQCP. Comparing to some previously
studied TQCPs \cite{Goswami11, Yang14B, Isobe16, Cho16}, this
topological transition exhibits richer critical phenomena and
provides interesting insight on the strong correlation effects at
the topology-changing phase transitions.

Recent theoretical studies \cite{Yang14A, LiuZunger17, Yu18}
predicted the presence of 3D cubic DSM state, in which the fermions
have the same dispersion as triple-Weyl fermions. When a 3D cubic
DSM is tuned to become a BI, the TQCP would exhibit the unusual
properties predicted in this paper. It is proposed that 3D cubic-DSM
can be realized in Rb(MoTe)$_{3}$ and Ti(MoTe)$_{3}$
\cite{LiuZunger17}, and LiOsO$_{3}$ \cite{Yu18}. More candidate
materials of triple-WSM or cubic-DSM might be discovered in the
future. If such materials are prepared with high quality, one can
tune them to approach the TQCP. We expect experiments would be
performed to measure the observable quantities studied in this paper
and to determine the nature of the low-energy instability. Quantum
Monte Carlo simulations \cite{Drut09, Ulybyshev13, Tupitsyn17,
Tang18} may help address these issues.

\emph{Note Added} - After we received the referees' reports, we had
become aware of two preprints \cite{Han18, Zhang18}, which study the
influence of Coulomb interaction at the TQCP between double-WSM and
BI.

\section*{ACKNOWLEDGEMENTS}

We acknowledge the support from the National Key R\&D Program of
China under Grants 2016YFA0300404 and 2017YFA0403600, and that from
the National Natural Science Foundation of China under Grants
11574285, 11504379, 11674327, U1532267, and U1832209.

\appendix

\section{Self-energy of Fermion \label{App:FermionSelfEnergy}}

The fermion self-energy is defined as
\begin{eqnarray}
\Sigma(i\omega,\mathbf{k}) &=& g^{2}\int'\frac{d\Omega}{2\pi}
\frac{d^3\mathbf{q}}{(2\pi)^{3}}G_{0}
\left(i\omega+i\Omega,\mathbf{k}+\mathbf{q}\right)\nonumber
\\
&&\times D_{0}(i\Omega,\mathbf{q}), \label{Eq:FermionSelfEnergy}
\end{eqnarray}
where the free fermion propagator is
\begin{eqnarray}
G_{0}(i\omega,\mathbf{k}) = \frac{1}{i\omega-\mathcal{H}_{f}(\mathbf{k})},
\label{Eq:FermionPropagator}
\end{eqnarray}
where
\begin{eqnarray}
\mathcal{H}_{f}(\mathbf{k}) = B[d_{1}(\mathbf{k}) \sigma_{1} +
d_{2}(\mathbf{k})\sigma_{2}]+Ad_{3}(\mathbf{k})\sigma_{3},
\end{eqnarray}
with $d_{1}(\mathbf{k})=\left(k_{x}^{3} - 3k_{x}k_{y}^{2}\right)$,
$d_{2}(\mathbf{k})=\left(k_{y}^{3} - 3k_{y}k_{x}^{2}\right)$, and
$d_{3}(\mathbf{k})=k_{z}^{2}$. The free boson propagator is given by
\begin{eqnarray}
D_{0}(i\Omega,\mathbf{q}) = D_{0}(\mathbf{q}) = \frac{1}{q_{x}^{2} +
q_{y}^{2} + \eta q_{z}^{2}} = \frac{1}{q_{\bot}^{2}+\eta q_{z}^{2}}.
\label{Eq:BosonPropagator}
\end{eqnarray}
Substituting Eqs.~(\ref{Eq:FermionPropagator}) and
(\ref{Eq:BosonPropagator}) into Eq.~(\ref{Eq:FermionSelfEnergy}),
and retaining the leading order of contribution, we obtain
\begin{eqnarray}
\Sigma(i\omega,\mathbf{k}) &\approx& i\omega\Sigma_{1} - B\left[
d_{1}(\mathbf{k})\sigma_{1} + d_{2}(\mathbf{k})\sigma_{2}\right]
\Sigma_{2} \nonumber \\
&&-Ad_{3}(\mathbf{k})\sigma_{3}\Sigma_{3},
\end{eqnarray}
where
\begin{eqnarray}
\Sigma_{1} &=& \frac{g^{2}}{4\pi^{3}}\int'd\Omega dq_{\bot}
d|q_{z}|q_{\bot} \frac{\Omega^{2} - B^{2}q_{\bot}^{6} - A^2
q_{z}^{4}}{\left(\Omega^{2}+B^{2}q_{\bot}^{6} + A^2
q_{z}^{4}\right)^{2}} \nonumber
\\
&&\times D_{0}(i\Omega,\mathbf{q}),
\\
\Sigma_{2}&=&\frac{g^{2}}{4\pi^{3}}\int'd\Omega
dq_{\bot}d|q_{z}|q_{\bot} \left[\frac{\Omega^{2}-18B^{2}q_{\bot}^{6}
+ A^2 q_{z}^{4}}{\left(\Omega^{2}+B^{2}q_{\bot}^{6} + A^2
q_{z}^{4}\right)^{2}} \right.\nonumber
\\
&&+ \frac{45B^{4}
q_{\bot}^{12}}{\left(\Omega^{2}+B^{2}q_{\bot}^{6} + A^2
q_{z}^{4}\right)^{3}}\nonumber
\\
&&\left. -\frac{27B^{6}q_{\bot}^{18}}{\left(\Omega^{2} + B^{2}
q_{\bot}^{6}+A^2q_{z}^{4}\right)^{4}}\right]D_{0}(i\Omega,\mathbf{q}),
\\
\Sigma_{3}&=&\frac{g^{2}}{4\pi^{3}}\int'd\Omega dq_{\bot}
d|q_{z}|q_{\bot} \left[\frac{\Omega^{2} + B^{2}
q_{\bot}^{6}-13A^{2}q_{z}^{4}}{\left(\Omega^{2} + B^{2}
q_{\bot}^{6}+A^2q_{z}^{4}\right)^{2}}\right.\nonumber
\\
&&\left.+\frac{16A^{4}q_{z}^{8}}{\left(\Omega^{2}+B^{2}q_{\bot}^{6} +
A^2q_{z}^{4}\right)^{3}}\right] D_{0}(i\Omega,\mathbf{q}).
\end{eqnarray}
In the following, we will employ the transformations
\begin{eqnarray}
E = \sqrt{B^2q_{\bot}^{6}+A^{2}q_{z}^{4}},\qquad \xi =
\frac{Bq_{\bot}^{3}}{A\left|q_{z}\right|^{2}},
\label{Eq:TransformationsA}
\end{eqnarray}
which are equivalent to
\begin{eqnarray}
q_{\bot}=\frac{\xi^{1/3} E^{1/3}}{B^{1/3}
\left(1+\xi^{2}\right)^{1/6}},\quad
|q_{z}|=\frac{\sqrt{E}}{\sqrt{A}\left(1+\xi^{2}\right)^{1/4}}.
\label{Eq:TranformationsB}
\end{eqnarray}
The measures of the integrations satisfy the relation
\begin{eqnarray}
dq_{\bot}d|q_{z}|&=& \left|\left|
\begin{array}{cc}
\frac{\partial q_{\bot}}{\partial E} & \frac{\partial
q_{\bot}}{\partial \xi} \\
\frac{\partial |q_{z}|}{\partial E} & \frac{\partial
|q_{z}|}{\partial \xi}
\end{array}
\right|\right|dEd\xi \nonumber
\\
&=& \left|\frac{\partial q_{\bot}}{\partial
E}\frac{\partial |q_{z}|}{\partial \xi} -\frac{\partial
q_{\bot}}{\partial \xi}\frac{\partial |q_{z}|}{\partial
E}\right|dEd\xi\nonumber
\\
&=& \frac{dEd\xi}{6\sqrt{A}B^{1/3}
E^{1/6} \xi^{2/3}
\left(1+\xi^{2}\right)^{5/12}}.
\label{Eq:TransformationsC}
\end{eqnarray}
To proceed with RG calculations, we choose to integrate out
high-energy modes defined in the momentum shell
\begin{eqnarray}
-\infty < \Omega < \infty, \qquad b\Lambda<E<\Lambda,
\label{Eq:RGScheme}
\end{eqnarray}
with $E= \sqrt{B^2q_{\bot}^{6}+A^2q_{z}^{4}}$ and $b = e^{-\ell}$.
Utilizing the transformations
Eqs.~(\ref{Eq:TransformationsA})-(\ref{Eq:TransformationsC}), we
obtain
\begin{eqnarray}
\Sigma(i\omega,\mathbf{k})&\approx& -B\left[d_{1}(\mathbf{k})\sigma_{1}
+d_{2}(\mathbf{k})\sigma_{2}\right] C_{2}\ell\nonumber
\\
&&-Ad_{3}(\mathbf{k})\sigma_{3}C_{3}\ell,
\end{eqnarray}
where
\begin{eqnarray}
C_{2}&=&\frac{g^{2}}{48\pi^{2}\sqrt{A\Lambda}}\int_{0}^{+\infty}
d\xi \frac{1}{\xi^{1/3}\left(1 +
\xi^{2}\right)^{13/12}}\nonumber
\\
&&\times \left[2-17\xi^{2}+\frac{135}{4}
\frac{\xi^{4}}{\left(1+\xi^{2}\right)} - \frac{135}{8}
\frac{\xi^{6}}{\left(1+\xi^{2}\right)^{2}}\right]\nonumber
\\
&&\times \frac{1}{\xi^{2/3}\left(1+\xi^{2}\right)^{1/6}
+ \zeta},
\\
C_{3}&=&\frac{g^{2}}{24\pi^{2}\sqrt{A\Lambda}}
\int_{0}^{+\infty}d\xi\frac{1}{\xi^{1/3}
\left(1+\xi^{2}\right)^{13/12}}\nonumber
\\
&&\times \left(\xi^{2} - 6 +
\frac{6}{1+\xi^{2}}\right)
\frac{1}{\xi^{2/3}
\left(1+\xi^{2}\right)^{1/6}+\zeta},
\end{eqnarray}
with $\zeta=\frac{\eta B^{2/3}\Lambda^{1/3}}{A}$.

\section{Self-energy of Boson \label{App:BosonSelfEnergy}}

The boson self-energy is given by
\begin{eqnarray}
\Pi(i\Omega,\mathbf{q})&=&-Ng^{2}\int\frac{d\omega}{2\pi}
\int'\frac{d^3\mathbf{k}}{(2\pi)^{3}}
\mathrm{Tr}\left[G_{0}(i\omega,\mathbf{k})\right.\nonumber
\\
&&\left.\times G_{0}\left(i\omega+i\Omega,\mathbf{k}+\mathbf{q}\right)\right].
\label{Eq:BonsonSelfEnergy}
\end{eqnarray}
Substituting Eq.~(\ref{Eq:FermionPropagator}) into
Eq.~(\ref{Eq:BonsonSelfEnergy}), $\Pi(i\Omega,\mathbf{q})$ can be
further written as
\begin{eqnarray}
\Pi(i\Omega,\mathbf{q}) &=& 2Ng^{2}\int_{-\infty}^{+\infty}
\frac{d\omega}{2\pi}\int'\frac{d^3\mathbf{k}}{(2\pi)^{3}}\nonumber
\\
&&\times\frac{1}
{\left(\omega^{2}+E_{\mathbf{k}}^{2}\right)
\left[(\omega+\Omega)^{2}+E_{\mathbf{k}+\mathbf{q}}^{2}\right]}
\big\{\omega(\omega+\Omega)\nonumber
\\
&&-B^{2}\left[d_{1}(\mathbf{k})
d_{1}(\mathbf{k}+\mathbf{q})+d_{2}(\mathbf{k})
d_{2}(\mathbf{k}+\mathbf{q})\right]\nonumber
\\
&& -A^2d_{3}(\mathbf{k}) d_{3}(\mathbf{k}+\mathbf{q})\big\},
\end{eqnarray}
where $E_{\mathbf{k}}=\sqrt{B^{2}k_{\bot}^{6}+A^{2}k_{z}^{4}}$.
Taking the static limit $\Omega=0$, and then expanding to the
leading order of $q_{\bot}$ and $q_{z}$, we get
\begin{eqnarray}
\Pi(0,\mathbf{q}) &\approx& q_{\bot}^{2} N\frac{9g^{2}}{16\pi^{2}}
\int'dk_{\bot}d|k_{z}|k_{\bot}\left(\frac{2B^{2}
k_{\bot}^{4}}{E_{\mathbf{k}}^{3}} - \frac{B^{4}
k_{\bot}^{10}}{E_{\mathbf{k}}^{5}}\right)\nonumber
\\
&& +q_{z}^{2}N \frac{g^{2}}{2\pi^{2}}\int'dk_{\bot}d|k_{z}|k_{\bot}
\frac{A^2k_{z}^{2} B^2k_{\bot}^{6}}{E_{\mathbf{k}}^{5}}.
\end{eqnarray}
Using again the transformations
Eqs.~(\ref{Eq:TransformationsA})-(\ref{Eq:TransformationsC}), we
finally obtain
\begin{eqnarray}
\Pi(0,\mathbf{q}) &\approx& q_{\bot}^{2}NC_{\bot}\ell +
q_{z}^{2}NC_{z}\ell, \nonumber
\end{eqnarray}
where
\begin{eqnarray}
C_{\bot} = \frac{9g^{2}}{40\pi^{2}\sqrt{A\Lambda}}, \quad C_{z} =
\frac{\Gamma\left(\frac{3}{4}\right)\Gamma\left(\frac{4}{3}\right)
g^{2}\sqrt{A}}{24\pi^{2}\Gamma\left(\frac{25}{12}\right)
B^{2/3}\Lambda^{5/6}}.
\end{eqnarray}

\section{Derivation of RG equations \label{App:DerivationRGQuation}}

The free action of fermion field $\psi$ is
\begin{eqnarray}
S_{\psi} &=& \int\frac{d\omega}{2\pi}\frac{d^{3}\mathbf{k}}{(2
\pi)^{3}} \psi_{a}^{\dag}(\omega,\mathbf{k})\left\{i\omega -
B[d_{1}(\mathbf{k})\sigma_{1}+d_{2}(\mathbf{k})\sigma_{2}]\right.\nonumber
\\
&&\left.-Ad_{3}(\mathbf{k})\sigma_{3}\right\}\psi_{a}(\omega,\mathbf{k}).
\end{eqnarray}
Upon incorporating the interaction corrections, we get
\begin{eqnarray}
S_{\psi} &=& \int\frac{d\omega}{2\pi} \frac{d^{3}
\mathbf{k}}{(2\pi)^{3}} \psi_{a}^{\dag}(\omega,\mathbf{k})
\left\{i\omega - B[d_{1}(\mathbf{k})\sigma_{1} + d_{2}(\mathbf{k})
\sigma_{2}]\right.\nonumber
\\
&&\left.-Ad_{3}(\mathbf{k})\sigma_{3} +
\Sigma(i\omega,\mathbf{k})\right\}\psi_{a}(\omega,\mathbf{k})
\nonumber \\
&\approx&\int\frac{d\omega}{2\pi}\frac{d^{3}\mathbf{k}}{(2\pi)^{3}}
\psi_{a}^{\dag}(\omega,\mathbf{k})\Big\{i\omega - B\left[
d_{1}(\mathbf{k})\sigma_{1}+d_{2}(\mathbf{k})\sigma_{2}\right]
\nonumber \\
&&\times e^{C_{2}\ell} - Ad_{3}(\mathbf{k}) \sigma_{3}
e^{C_{3}\ell}\Big\}\psi_{a}(\omega,\mathbf{k}).
\end{eqnarray}
Using the re-scaling transformations
\begin{eqnarray}
\omega&=&\omega'e^{-\ell},\label{Eq:ScalingOmega}
\\
k_{x}&=&k_{x}'e^{-\frac{\ell}{3}}, \label{Eq:Scalingkx}
\\
k_{y}&=&k_{y}'e^{-\frac{\ell}{3}}, \label{Eq:Scalingky}
\\
k_{z}&=&k_{z}'e^{-\frac{\ell}{2}}, \label{Eq:Scalingkz}
\\
\psi_{a}&=&\psi_{a}' e^{\frac{19}{12}\ell}, \label{Eq:Scalingpsi}
\\
B&=&B'e^{-C_{2}\ell}, \label{Eq:ScalingB}
\\
A&=&A'e^{-C_{3}\ell}, \label{Eq:ScalingA}
\end{eqnarray}
the action can be re-written as
\begin{eqnarray}
S_{\psi'}&=&\int\frac{d\omega'}{2\pi}\frac{d^{3}\mathbf{k}'}{(2
\pi)^{3}}{\psi'}_{a}^{\dag}(\omega',\mathbf{k'}) \big\{i\omega'
- B'\left[d_{1}(\mathbf{k}')\sigma_{1}\right. \nonumber \\
&&\left. + d_{2}(\mathbf{k}')\sigma_{2}\right] -
A'd_{3}(\mathbf{k}')\sigma_{3}\big\}\psi_{a}'(\omega',\mathbf{k}'),
\end{eqnarray}
which recovers the original form of the action.

The free action of boson field $\phi$ takes the form
\begin{eqnarray}
S_{\phi}=\int\frac{d\omega}{2\pi}\frac{d^{3}\mathbf{k}}{(2\pi)^{3}}
\phi(\omega,\mathbf{k}) \left(k_{\bot}^{2}+\eta k_{z}^{2}
\right)\phi(\omega,\mathbf{k}).
\end{eqnarray}
Including the interaction correction converts the action $\phi$ to
\begin{eqnarray}
S_{\phi} &=& \int\frac{d\omega}{2\pi} \frac{d^{3}
\mathbf{k}}{(2\pi)^{3}} \phi(\omega,\mathbf{k})
\left[k_{\bot}^{2}+\eta k_{z}^{2} + \Pi(0,\mathbf{k})\right]
\phi(\omega,\mathbf{k})\nonumber \\
&\approx&\int\frac{d\omega}{2\pi}\frac{d^{3}\mathbf{k}}{(2\pi)^{3}}
\phi(\omega,\mathbf{k})\left[k_{\bot}^{2}e^{NC_{\bot}\ell} +
\left(\eta+NC_{z}\ell\right)k_{z}^{2}\right]\nonumber
\\
&&\times\phi(\omega,\mathbf{k}).
\end{eqnarray}
Making use of the transformations
Eqs.~(\ref{Eq:ScalingOmega})-(\ref{Eq:Scalingkz}), we find that the
field $\phi$ should be re-scaled as follows
\begin{eqnarray}
\phi = \phi'e^{-\frac{1}{2}\left(-\frac{17}{6}+NC_{\bot}\right)\ell}.
\label{Eq:Scalingphi}
\end{eqnarray}
Now the action of $\phi'$ can be approximately written as
\begin{eqnarray}
S_{\phi'} &\approx&
\int\frac{d\omega'}{2\pi}\frac{d^{3}\mathbf{k}'}{(2\pi)^{3}}
\phi'(\omega',\mathbf{k}')
\bigg\{k_{\bot}'^{2}+\Big[\eta-\eta\Big(\frac{1}{3} + NC_{\bot}
\Big) \nonumber \\
&&\times\ell + NC_{z}\ell\Big]k_{z}'^{2}\bigg\}
\phi'(\omega',\mathbf{k}').
\end{eqnarray}
We define
\begin{eqnarray}
\eta' = \eta-\eta\left(\frac{1}{3} +
NC_{\bot}\right)\ell+NC_{z}\ell, \label{Eq:Scalingeta}
\end{eqnarray}
and then get
\begin{eqnarray}
S_{\phi'} &=& \int\frac{d\omega'}{2\pi}
\frac{d^{3}\mathbf{k}'}{(2\pi)^{3}}
\phi'(\omega',\mathbf{k}')
\left(k_{\bot}'^{2}+\eta'k_{z}'^{2}\right)\nonumber
\\
&&\times\phi'(\omega',\mathbf{k}'),
\end{eqnarray}
which has the same form as the original boson action.

\begin{figure*}[htbp]
\center
\includegraphics[width=6.3in]{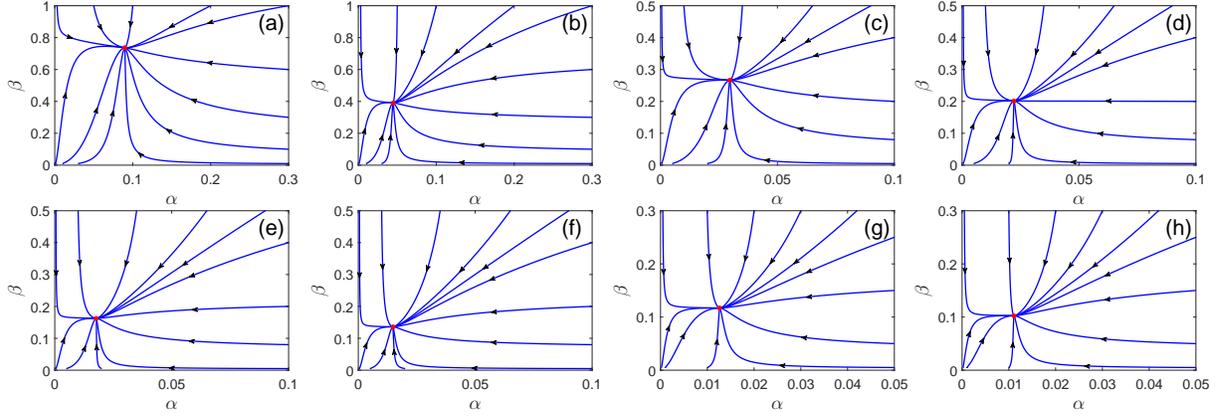}
\caption{(a)-(h): Flow diagrams on the $\alpha$-$\beta$ plane at
$N=1,2,3,4,5,6,7,8$, respectively. \label{Fig:FlowDiagramNoFFSuMt}}
\end{figure*}

The action of fermion-boson coupling is
\begin{eqnarray}
S_{\psi\phi} &=& g\int\frac{d\omega_{1}}{2\pi}
\frac{d^3\mathbf{k}_{1}}{(2\pi)^{3}}\frac{d\omega_{2}}{2\pi}
\frac{d^3\mathbf{k}_{2}}{(2\pi)^{3}}
\psi_{a}^{\dag}(\omega_{1},\mathbf{k}_{1})
\psi_{a}(\omega_{2},\mathbf{k}_{2})\nonumber
\\
&&\times\phi(\omega_{1}-\omega_{2},\mathbf{k}_{1}
- \mathbf{k}_{2}).
\end{eqnarray}
Making use of the transformations
Eqs.~(\ref{Eq:ScalingOmega})-(\ref{Eq:Scalingpsi}) and
(\ref{Eq:Scalingphi}), we find that
\begin{eqnarray}
S_{\psi'\phi'} &=& ge^{\left(\frac{1}{4} -
\frac{NC_{\bot}}{2}\right)\ell} \int\frac{d\omega_{1}'}{2\pi}
\frac{d^3\mathbf{k}_{1}'}{(2\pi)^{3}}\frac{d\omega_{2}'}{2\pi}
\frac{d^3\mathbf{k}_{2}'}{(2\pi)^{3}}
\psi'^{\dag}_{a}(\omega_{1}',\mathbf{k}_{1}')\nonumber
\\
&&\times\psi'_{a}(\omega_{2}',\mathbf{k}_{2}')
\phi'(\omega_{1}'-\omega_{2}',\mathbf{k}_{1}'-\mathbf{k}_{2}).
\end{eqnarray}
Adopting the re-scaling transformation
\begin{eqnarray}
g' = ge^{\left(\frac{1}{4} - \frac{NC_{\bot}}{2}\right)\ell},
\label{Eq:gScaling}
\end{eqnarray}
we further re-write the action in the form
\begin{eqnarray}
S_{\psi'\phi'} &=& g'\int\frac{d\omega_{1}'}{2\pi}
\frac{d^3\mathbf{k}_{1}'}{(2\pi)^{3}}\frac{d\omega_{2}'}{2\pi}
\frac{d^3\mathbf{k}_{2}'}{(2\pi)^{3}}
\psi'^{\dag}_{a}(\omega_{1}',\mathbf{k}_{1}')\nonumber
\\
&&\times\psi'_{a}(\omega_{2}',\mathbf{k}_{2}')
\phi'_{a}(\omega_{1}'-\omega_{2}',\mathbf{k}_{1}' -\mathbf{k}_{2}'),
\end{eqnarray}
which recovers the form of the original interacting action.

From the transformations given by
Eqs.~(\ref{Eq:ScalingB}), (\ref{Eq:ScalingA}),
(\ref{Eq:Scalingeta}), and (\ref{Eq:gScaling}), we obtain the
following RG equations
\begin{eqnarray}
\frac{dB}{d\ell}&=&C_{2}B,
\\
\frac{dA}{d\ell}&=&C_{3}A,
\\
\frac{d\alpha}{d\ell} &=& \left(\frac{1}{2}-NC_{\bot} -
\frac{C_{3}}{2}\right)\alpha,
\\
\frac{d\beta}{d\ell} &=& \left(\frac{5}{6} +\frac{1}{2}C_{3}
-\frac{2}{3}C_{2}-N\beta\right)\beta,
\\
\frac{d\eta}{d\ell} &=&
\left(-\frac{1}{3}-NC_{\bot}+N\beta\right)\eta,
\\
\frac{d\zeta}{d\ell} &=& \left(-\frac{1}{3} + \frac{2}{3}C_{2} -
C_{3} - N C_{\bot}+N\beta\right)\zeta,
\end{eqnarray}
where $\alpha$ and $\beta$ are defined as
\begin{eqnarray}
\alpha&=&\frac{g^{2}}{8\pi^{3} \sqrt{A\Lambda}},
\\
\beta&=&\frac{C_{z}}{\eta}=\frac{\Gamma\left(\frac{3}{4}\right)
\Gamma\left(\frac{4}{3}\right)g^{2}\sqrt{A}}{24\pi^{2}
\Gamma\left(\frac{25}{12}\right)B^{2/3}
\Lambda^{5/6}\eta}.
\end{eqnarray}

As shown in Fig.~\ref{Fig:FlowDiagramNoFFSuMt}, ($\alpha$,$\beta$)
always flows to a stable fixed point $(\alpha^{*},\beta^{*})$. The
concrete values of $\alpha^{*}$ and $\beta^{*}$ are determined by
the fermion flavor $N$. The values of $\alpha^{*}$, $\beta^{*}$, and
several other parameters obtained for different fermion flavors are
summarized in Table~\ref{Table:FixedPointParameters}.

\begin{table*}[htbp]
\caption{Dependence of the parameters $\alpha^{*}$, $\beta^{*}$,
$C_{2}^{*}$, $C_{3}^{*}$, $C_{\bot}^{*}$, $N\alpha^{*}$,
$N\beta^{*}$, $NC_{\bot}^{*}$ on $N$ at the stable fixed point.
\label{Table:FixedPointParameters}}
\begin{center}
\begin{tabular}{|c|c|c|c|c|c|c|c|c|}
\hline\hline    $N$ & $\alpha^{*}$ & $\beta^{*}$ & $C_{2}^{*}$ &
$C_{3}^{*}$ & $C_{\bot}^{*}$ & $N\alpha^{*}$ & $N\beta^{*}$ &  $N
C_{\bot}^{*}$
\\
\hline          1& 0.08997 & 0.7364  & 0.1322 & -0.01754 &  0.5088 &
0.08997 & 0.7364  & 0.5088
\\
\hline          2& 0.04462 & 0.3914  & 0.06891 & -0.009236 &  0.2523
& 0.08924 & 0.7828  & 0.5046
\\
\hline          3& 0.02966 & 0.2664 & 0.04567 & -0.006258 &  0.1677
& 0.08897 & 0.7992  & 0.5031
\\
\hline          4& 0.02221 & 0.2019 & 0.03517 & -0.004731 &  0.1256
& 0.08884 & 0.8075  & 0.5024
\\
\hline          5& 0.01775 & 0.1625 & 0.02825 & -0.003802 &  0.1004
& 0.08876 & 0.8126  & 0.5019
\\
\hline          6& 0.01478 & 0.1360 & 0.02361 & -0.003178 &  0.08360
& 0.08870 & 0.8160  & 0.5016
\\
\hline          7& 0.01267 & 0.1169 & 0.02027 & -0.002730 &  0.07162
& 0.08866 & 0.8185  & 0.5014
\\
\hline          8& 0.01108 & 0.1025 & 0.01776 & -0.002393 &  0.06265
& 0.08863 & 0.8203  & 0.5012
\\
\hline          9& 0.009845 & 0.09130 & 0.01581 & -0.002130 &
0.05567 & 0.08861 & 0.8217  & 0.5010
\\
\hline          10& 0.008859 & 0.08229 & 0.01424 & -0.001919 &
0.05010 & 0.08859 & 0.8229  & 0.5010
\\
\hline \hline
\end{tabular}
\end{center}
\end{table*}

\section{Observable quantities in the free case \label{App:ObserQuanFree}}

Here, we will calculate a number of observable quantities, first in
the non-interacting limit.

\subsection{Density of states}

The retarded fermion propagator takes the form
\begin{eqnarray}
G_{0}^{\mathrm{ret}}(\omega,\mathbf{k}) = \frac{1}{\omega
-\mathcal{H}_{f}(\mathbf{k})+i\delta}.
\end{eqnarray}
Its imaginary part can be easily obtained:
\begin{eqnarray}
\mathrm{Im}\left[G_{0}^{\mathrm{ret}}(\omega,\mathbf{k})\right] &=&
-\pi\mathrm{sgn}(\omega)\left[\omega+\mathcal{H}_{f}(\mathbf{k})\right]
\frac{1}{2E_{\mathbf{k}}}\nonumber
\\
&&\times\left[\delta
\left(\omega+E_{\mathbf{k}}\right) + \delta
\left(\omega-E_{\mathbf{k}}\right)\right].
\label{Eq:ImPartRetFermionPropagator}
\end{eqnarray}
The spectral function is given by
\begin{eqnarray}
\mathcal{A}(\omega,\mathbf{k}) &=& -\frac{1}{\pi}
\mathrm{Tr}\left[\mathrm{Im}
\left[G_{0}^{\mathrm{ret}}(\omega,\mathbf{k})\right]\right]\nonumber
\\
&=&\left|\omega\right|\frac{1}{E_{\mathbf{k}}}\left[\delta
\left(\omega+E_{\mathbf{k}}\right) + \delta
\left(\omega-E_{\mathbf{k}}\right)\right],
\end{eqnarray}
which directly leads to the fermion DOS
\begin{eqnarray}
\rho(\omega) &=& N \int\frac{d^3\mathbf{k}}{(2\pi)^{3}}
\mathcal{A}(\omega,\mathbf{k})\nonumber
\\
&=& \frac{|\omega|}{2\pi^{2}}\int
dk_{\bot}d|k_{z}|k_{\bot} \frac{1}{E_{\mathbf{k}}}\left[\delta
\left(\omega+E_{\mathbf{k}}\right)\right.\nonumber
\\
&&\left. + \delta
\left(\omega-E_{\mathbf{k}}\right)\right].
\end{eqnarray}
After integrating over the momenta, we obtain
\begin{eqnarray}
\rho(\omega) = \frac{Nc_{1}}{
\sqrt{A}B^{2/3}}|\omega|^{1/6},
\end{eqnarray}
where
\begin{eqnarray}
c_{1} &=& \frac{\Gamma\left(\frac{1}{3}\right)
\Gamma\left(\frac{5}{4}\right)}{6\pi^{2}
\Gamma\left(\frac{7}{12}\right)}\approx 0.02682.
\end{eqnarray}

\subsection{Specific heat}

The Matsubara fermion propagator has the form
\begin{eqnarray}
G_{0}(i\omega_{n},\mathbf{k}) &=& \frac{1}{i\omega_{n} -
\mathcal{H}_{f}(\mathbf{k})}\nonumber \\
&=&\frac{-i\omega_{n}-\mathcal{H}_{f}(\mathbf{k})}{\omega_{n}^{2} +
E_{\mathbf{k}}^{2}}, \label{Eq:FermionPropagatorFinitT}
\end{eqnarray}
where $\omega_{n}=(2n+1)\pi T$ with $n$ being integers. The free
energy of fermions is
\begin{eqnarray}
F_{f}(T) &=& -2NT\sum_{\omega_{n}} \int\frac{d^3
\mathbf{k}}{(2\pi)^3} \ln\left[\left(\omega_{n}^2 +
E_{\mathbf{k}}^{2}\right)^{1/2}\right].
\end{eqnarray}
Performing frequency summation yields
\begin{eqnarray}
F_{f}(T) = -2N\int\frac{d^3\mathbf{k}}{(2\pi)^3} \left[E_{\mathbf{k}}
+ 2T\ln\left(1+e^{-\frac{E_{\mathbf{k}}}{T}}\right)\right],
\end{eqnarray}
which is divergent due to the first term in bracket. To get a finite
free energy, we replace $F_{f}(T)$ with $F_{f}(T)-F_{f}(0)$, and
thus get
\begin{eqnarray}
F_{f}(T) &=& -4NT\int\frac{d^3\mathbf{k}}{(2\pi)^3}
\ln\left(1+e^{-\frac{E_{\mathbf{k}}}{T}}\right)\nonumber
\\
&=&-\frac{2NT}{\pi^{2}}\int dk_{\bot} d|k_{z}| k_{\bot}
\ln\left(1+e^{-\frac{E_{\mathbf{k}}}{T}}\right).
\end{eqnarray}
Employing the transformations
Eqs.~(\ref{Eq:TransformationsA})-(\ref{Eq:TransformationsC}) and
then carrying out the integrations, we have
\begin{eqnarray}
F_{f}(T) &= &-\frac{\left(4-2^{5/6}\right)
\zeta\left(\frac{13}{6}\right)\Gamma\left(\frac{7}{6}\right)
\Gamma\left(\frac{1}{3}\right)
\Gamma\left(\frac{5}{4}\right)N}{6\pi^{2}
\Gamma\left(\frac{7}{12}\right)\sqrt{A}B^{2/3}}\nonumber
\\
&&\times T^{13/6},
\end{eqnarray}
where $\zeta(x)$ is the Riemann zeta function. The corresponding
specific heat satisfies
\begin{eqnarray}
C_{v}(T) = -T\frac{\partial^2 F_{f}(T)}{\partial T^2} =
\frac{Nc_{2}}{\sqrt{A}B^{2/3}}T^{7/6},
\end{eqnarray}
where
\begin{eqnarray}
c_{2} &=& \frac{91\left(4-2^{5/6}\right)
\zeta\left(\frac{13}{6}\right)\Gamma\left(\frac{7}{6}\right)
\Gamma\left(\frac{1}{3}\right)\Gamma\left(\frac{5}{4}\right)}{216
\pi^{2}\Gamma\left(\frac{7}{12}\right)} \nonumber
\\
&\approx& 0.211.
\end{eqnarray}

\subsection{Compressibility}

In order to calculate the compressibility, we formally include a
finite chemical potential $\mu$ into the fermion propagator, namely
\begin{eqnarray}
G_{0}(i\omega_{n},\mathbf{k}) = \frac{-i\omega_{n} - \mu -
\mathcal{H}_{f}(\mathbf{k})}{\left(\omega_{n}-i\mu\right)^{2}+E_{\mathbf{k}}^{2}}.
\end{eqnarray}
The corresponding free energy now becomes
\begin{eqnarray}
F_{f}(T,\mu) &=& -2NT\sum_{\omega_{n}} \int
\frac{d^3\mathbf{k}}{(2\pi)^3}\nonumber
\\
&&\times\ln\left[\left(\left(\omega_{n} -
i\mu\right)^2 + E_{\mathbf{k}}^{2}\right)^{1/2}\right].
\end{eqnarray}
Summing over frequencies allows us to get
\begin{eqnarray}
F_{f}(T,\mu) &=& -2NT\int\frac{d^3\mathbf{k}}{(2\pi)^3}
\left[\ln\left(1+e^{-\frac{E_{\mathbf{k}}-\mu}{T}}\right)\right.\nonumber
\\
&&\left. +
\ln\left(1+e^{-\frac{E_{\mathbf{k}}+\mu}{T}}\right)\right]\nonumber
\\
&=&-\frac{NT}{\pi^{2}}\int dk_{\bot} d|k_{z}| k_{\bot}
\left[\ln\left(1+e^{-\frac{E_{\mathbf{k}}-\mu}{T}}\right)\right.\nonumber
\\
&&\left.+\ln\left(1+e^{-\frac{E_{\mathbf{k}}+\mu}{T}}\right)\right].
\end{eqnarray}
Using the transformations
Eqs.~(\ref{Eq:TransformationsA})-(\ref{Eq:TransformationsC}) and
carrying out integration over momenta, we obtain
\begin{eqnarray}
F_{f}(T,\mu) &=& \frac{N\Gamma\left(\frac{1}{3}\right)
\left(\frac{5}{4}\right)\Gamma\left(\frac{7}{6}\right)}{3\pi^{2}
\Gamma\left(\frac{7}{12}\right)\sqrt{A}B^{2/3}} T^{13/6}
\left[\mathrm{Li}_{\frac{13}{6}}\left(-e^{\frac{\mu}{T}}\right)\right.\nonumber
\\
&&\left.+\mathrm{Li}_{\frac{13}{6}}\left(-e^{-\frac{\mu}{T}}\right)\right],
\end{eqnarray}
where $\mathrm{Li}_{x}(y)$ is the polylogarithmic function. The
compressibility is
\begin{eqnarray}
\kappa(T,\mu) &=& -\frac{\partial^2 F_{f}(T,\mu)}{\partial \mu^2}\nonumber
\\
&=&-\frac{N\Gamma\left(\frac{1}{3}\right)\left(\frac{5}{4}\right)
\Gamma\left(\frac{7}{6}\right)}{3\pi^{2}\Gamma(\frac{7}{12}) \sqrt{A}
B^{2/3}}T^{1/6} \left[\mathrm{Li}_{\frac{1}{6}}
\left(-e^{\frac{\mu}{T}}\right)\right.\nonumber
\\
&&\left. +\mathrm{Li}_{\frac{1}{6}}
\left(-e^{-\frac{\mu}{T}}\right)\right].
\end{eqnarray}
Taking the limit $\mu=0$, we finally get
\begin{eqnarray}
\kappa (T) = \frac{Nc_{3}} {\sqrt{A}B^{2/3}}T^{1/6},
\end{eqnarray}
where
\begin{eqnarray}
c_{3} &=& \frac{2\left(1-2^{5/6}\right)
\zeta\left(\frac{1}{6}\right)\Gamma\left(\frac{1}{3}\right) \Gamma
\left(\frac{5}{4}\right)\Gamma\left(\frac{7}{6}\right)}{3\pi^{2}
\Gamma\left(\frac{7}{12}\right)}\nonumber
\\
&\approx& 0.05343.
\end{eqnarray}

\subsection{Dynamical conductivities}

The dynamical conductivities can be obtained by calculating the
current-current correlation function, which in the Matsubara
formalism is usually defined as
\begin{eqnarray}
\Pi_{ij}(i\Omega_{m}) &=& -e^{2}T\sum_{\omega_{n}}\int
\frac{d^3\mathbf{k}}{(2\pi)^{3}}\mathrm{Tr}\left[\gamma_{i}(\mathbf{k})
G_{0}(i\omega_{n},\mathbf{k})\gamma_{j}(\mathbf{k})\right.\nonumber
\\
&&\left.\times G_{0}(i\omega_{n}+i\Omega_{m},\mathbf{k})\right],
\end{eqnarray}
where $\Omega_{m}=2m\pi T$ with $m$ being integers.
Here, $\gamma_{i}$ is given by
\begin{eqnarray}
\gamma_{i} = \frac{\partial \mathcal{H}_{f}}{\partial k_{i}}.
\end{eqnarray}
It is easy to verify that
\begin{eqnarray}
\gamma_{x} &=& \frac{\partial \mathcal{H}_{f}}{\partial k_{x}} =
3B\left(k_{x}^{2} - k_{y}^{2}\right)\sigma_{1} -
6Bk_{x}k_{y}\sigma_{2}, \label{Eq:gammaxDef} \\
\gamma_{y}&=&\frac{\partial \mathcal{H}_{f}}{\partial k_{y}} = -
6Bk_{x}k_{y} \sigma_{1} + 3B\left(k_{y}^{2} -
k_{x}^{2}\right)\sigma_{2}, \label{Eq:gammayDef}
\\
\gamma_{z} &=& \frac{\partial \mathcal{H}_{f}}{\partial k_{z}} =
2Ak_{z}\sigma_{3}. \label{Eq:gammazDef}
\end{eqnarray}
The symmetry of the Hamiltonian dictates that $$\Pi_{xx} =
\Pi_{yy}\equiv\Pi_{\bot\bot},$$ so it is only necessary to calculate
$\Pi_{xx}$ and $\Pi_{zz}$, defined as follows
\begin{eqnarray}
\Pi_{xx}(i\Omega_{m}) &=& -e^{2}T\sum_{\omega_{n}}\int
\frac{d^3\mathbf{k}}{(2\pi)^{3}}\mathrm{Tr}\left[\gamma_{x}
G_{0}(i\omega_{n},\mathbf{k})\gamma_{x}\right.\nonumber
\\
&&\left.\times G_{0}(i\omega_{n}+i\Omega_{m},\mathbf{k})\right],
\label{Eq:PolaxxDef}
\\
\Pi_{zz}(i\Omega_{m})&=&-e^{2}T\sum_{\omega_{n}}\int
\frac{d^3\mathbf{k}}{(2\pi)^{3}}\mathrm{Tr}\left[\gamma_{z}
G_{0}(i\omega_{n},\mathbf{k})\gamma_{z}\right.\nonumber
\\
&&\left.\times G_{0}(i\omega_{n}+i\Omega_{m},\mathbf{k})\right].
\label{Eq:PolazzDef}
\end{eqnarray}
With the help of spectral representation
\begin{equation}
G_0\left(i\omega_n,\mathbf{k}\right)=-\int_{-\infty}^{+\infty}
\frac{d\omega_1}{\pi}\frac{\mathrm{Im}\left[G_0^{\mathrm{ret}}
\left(\omega_1,\mathbf{k}\right)\right]}{i\omega_n-\omega_1},
\end{equation}
we now can re-write $\Pi_{xx}$ and $\Pi_{zz}$ as follows
\begin{widetext}
\begin{eqnarray}
\Pi_{xx}(i\Omega_{m}) &=& -9B^{2}e^{2}\int
\frac{d^3\mathbf{k}}{(2\pi)^{3}}\int_{-\infty}^{+\infty}
\frac{d\omega_1}{\pi}\int_{-\infty}^{+\infty}\frac{d\omega_2}{\pi}
\Big\{\left(k_{x}^{2}-k_{y}^{2}\right)^{2} \mathrm{Tr}
\left[\sigma_{1}\mathrm{Im}\left[G_0^{\mathrm{ret}}
\left(\omega_1,\mathbf{k}\right)\right]\sigma_{1}\mathrm{Im}
\left[G_0^{\mathrm{ret}}\left(\omega_2,\mathbf{k}
\right)\right]\right] \nonumber \\
&&+4k_{x}^{2}k_{y}^{2}\mathrm{Tr}\left[\sigma_{2} \mathrm{Im}
\left[G_0^{\mathrm{ret}}\left(\omega_1,\mathbf{k}\right)\right]
\sigma_{2} \mathrm{Im}\left[G_0^{\mathrm{ret}}
\left(\omega_2,\mathbf{k}\right)\right]\right]\Big\}
\frac{n_F\left(\omega_1\right)-n_F\left(\omega_2\right)}{\omega_1 -
\omega_2+i\Omega_{m}}, \label{Eq:PolaxxMidA}
\\
\Pi_{zz}(i\Omega_{m}) &=& -4A^{2}e^{2}\int
\frac{d^3\mathbf{k}}{(2\pi)^{3}} k_{z}^{2}\int_{-\infty}^{+\infty}
\frac{d\omega_1}{\pi}\int_{-\infty}^{+\infty} \frac{d\omega_2}{\pi}
\mathrm{Tr}\left[\sigma_{3}\mathrm{Im}\left[G_0^{\mathrm{ret}}
\left(\omega_1,\mathbf{k}\right)\right]\sigma_{3}
\mathrm{Im}\left[G_0^{\mathrm{ret}}
\left(\omega_2,\mathbf{k}\right)\right]\right]\nonumber
\\
&&\times \frac{n_F\left(\omega_1\right) -
n_F\left(\omega_2\right)}{\omega_1-\omega_2 + i\Omega_{m}},
\label{Eq:PolazzMidA}
\end{eqnarray}
where $n_{F}(x)=\frac{1}{e^{x/T}+1}$. We then carry out analytical
continuation $i\Omega_{m}\rightarrow \Omega + i\delta$, and get the
imaginary parts:
\begin{eqnarray}
\mathrm{Im}\left[\Pi_{xx}^{\mathrm{ret}}(\Omega, T)\right] &=&
9B^{2}e^{2}\int\frac{d^3\mathbf{k}}{(2\pi)^{3}}\int_{-\infty}^{+\infty}
\frac{d\omega_1}{\pi}\Big\{\left(k_{x}^{2}-k_{y}^{2}\right)^{2}
\mathrm{Tr}\left[\sigma_{1}\mathrm{Im}\left[G_0^{\mathrm{ret}}
\left(\omega_1,\mathbf{k}\right)\right]\sigma_{1}
\mathrm{Im}\left[G_0^{\mathrm{ret}} \left(\omega_1 +
\Omega,\mathbf{k}\right)\right]\right]\nonumber
\\
&&+4k_{x}^{2}k_{y}^{2}\mathrm{Tr}\left[\sigma_{2}
\mathrm{Im}\left[G_0^{\mathrm{ret}}
\left(\omega_1,\mathbf{k}\right)\right]\sigma_{2}
\mathrm{Im}\left[G_0^{\mathrm{ret}} \left(\omega_1 +
\Omega,\mathbf{k}\right)\right]\right]\Big\}
\left[n_F\left(\omega_1\right)-n_F\left(\omega_1+\Omega\right)\right],
\label{Eq:PolaxxMidB}
\\
\mathrm{Im}\left[\Pi_{zz}^{\mathrm{ret}}(\Omega,T)\right] &=&
4A^{2}e^{2}\int\frac{d^3\mathbf{k}}{(2\pi)^{3}}k_{z}^{2}
\int_{-\infty}^{+\infty} \frac{d\omega_1}{\pi}
\mathrm{Tr}\left[\sigma_{3}\mathrm{Im}\left[G_0^{\mathrm{ret}}
\left( \omega_1,\mathbf{k}\right)\right]\sigma_{3}
\mathrm{Im}\left[G_0^{\mathrm{ret}}\left(\omega_1 +
\Omega,\mathbf{k}\right)\right]\right]\nonumber
\\
&&\times\left[n_F\left(\omega_1\right) -
n_F\left(\omega_1+\Omega\right)\right]. \label{Eq:PolazzMidB}
\end{eqnarray}
\end{widetext}
The formula $\frac{1}{x+i\delta} = \mathcal{P} \frac{1}{x} -
i\pi\delta(x)$, where $\mathcal{P}\frac{1}{x}$ is the principal
value, was used above. The conductivities are related to
current-current correlation function as follows
\begin{eqnarray}
\sigma_{xx}(\Omega,T)&=&\frac{\mathrm{Im}
\left[\Pi_{xx}^{\mathrm{ret}}(\Omega,T)\right]}{\Omega},
\label{Eq:SigmaxxDef}
\\
\sigma_{zz}(\Omega,T)&=&\frac{\mathrm{Im}
\left[\Pi_{zz}^{\mathrm{ret}}(\Omega,T)\right]}{\Omega}.
\label{Eq:SigmazzDef}
\end{eqnarray}
After accomplishing tedious but straightforward calculations, we
eventually obtain
\begin{eqnarray}
\sigma_{xx}(\Omega,T) &=& \frac{3e^{2}}{4\pi\sqrt{A}}
\delta\left(\Omega\right)T^{3/2}\int_{0}^{+\infty}
dxx^{3/2}\frac{1}{\sinh^{2}\left(\frac{x}{2}\right)}\nonumber
\\
&&+\frac{c_{4}e^{2}}{\sqrt{A}}|\Omega|^{1/2}
\tanh\left(\frac{|\Omega|}{4T}\right), \label{Eq:SigmaxxResult}
\\
\sigma_{zz}(\Omega,T) &=&
\frac{36\Gamma\left(\frac{1}{3}\right)\Gamma\left(\frac{3}{4}\right)
\sqrt{A}e^{2}}{13\Gamma\left(\frac{1}{12}\right)\pi B^{2/3}}
\delta\left(\Omega\right)T^{7/6}\nonumber
\\
&&\times\int_{0}^{+\infty}
dxx^{7/6}\frac{1}{\sinh^{2}\left(\frac{x}{2}\right)}\nonumber
\\
&&+\frac{c_{5}\sqrt{A}e^{2}}{ B^{2/3}}|\Omega|^{1/6}
\tanh\left(\frac{|\Omega|}{4T}\right), \label{Eq:SigmazzResult}
\end{eqnarray}
where
\begin{eqnarray}
c_{4}&=&\frac{9}{8\sqrt{2}\pi}\approx0.2532, \nonumber
\\
c_{5}&=&\frac{2^{5/6}\Gamma\left(\frac{1}{3}\right)
\Gamma\left(\frac{3}{4}\right)}{13\Gamma\left(\frac{1}{12}\right)\pi}
\approx 0.01245.
\end{eqnarray}
The first terms in the right-hand side of
Eqs.~(\ref{Eq:SigmaxxResult}) and (\ref{Eq:SigmazzResult}) represent
the Drude peak.

\subsection{Diamagnetic Susceptibilities}

The diamagnetic susceptibility is defined as
\begin{eqnarray}
\chi_{D} &=& -Ne^{2}T\sum_{\omega_{n}}\int
\frac{d^3\mathbf{k}}{(2\pi)^{3}}\mathrm{Tr}\left[\gamma_{a}
G_{0}(i\omega_{n},\mathbf{k})\gamma_{b}
\right.\nonumber
\\
&&\left.\times G_{0}(i\omega_{n},\mathbf{k})\gamma_{a}G_{0}(i\omega_{n},\mathbf{k})
\gamma_{b}G_{0}(i\omega_{n},\mathbf{k})\right].
\end{eqnarray}
Here, $\gamma_{a}$ is given by
\begin{eqnarray}
\gamma_{a} = \frac{\partial \mathcal{H}_{f}}{\partial k_{a}},
\end{eqnarray}
where $a$ denotes an axis that is perpendicular to the direction of
external magnetic field. Owing to the rotational symmetry of the
system, the diamagnetic susceptibility should exhibit the same
behavior for fields aligned along $x$ and $y$ axes, i.e.,
$\chi_{D}^{x}=\chi_{D}^{y}\equiv \chi_{D}^{\bot}$. However, for
magnetic field $\mathbf{B} = B\mathbf{e}_{z}$, diamagnetic
susceptibility should be different. By taking
$\mathbf{B}=B\mathbf{e}_{x}$ as an example, we express the
diamagnetic susceptibility in the form
\begin{eqnarray}
\chi_{D}^{\bot} &=& -Ne^{2}T\sum_{\omega_{n}}\int
\frac{d^3\mathbf{k}}{(2\pi)^{3}} \mathrm{Tr}
\left[\gamma_{y}G_{0}(i\omega_{n},\mathbf{k})\gamma_{z}
G_{0}(i\omega_{n},\mathbf{k})\right.\nonumber
\\
&&\left.\times \gamma_{y}G_{0}(i\omega_{n},\mathbf{k})
\gamma_{z}G_{0}(i\omega_{n},\mathbf{k})\right].
\label{Eq:SusceptBotDef}
\end{eqnarray}
In the case of magnetic field $\mathbf{B} = B\mathbf{e}_{z}$, the
diamagnetic susceptibility is defined as
\begin{eqnarray}
\chi_{D}^{z} &=& -Ne^{2}T\sum_{\omega_{n}}\int \frac{d^3
\mathbf{k}}{(2\pi)^{3}} \mathrm{Tr}\left[\gamma_{x}
G_{0}(i\omega_{n},\mathbf{k})\gamma_{y}
G_{0}(i\omega_{n},\mathbf{k})\right.\nonumber
\\
&&\left.\times\gamma_{x}G_{0}(i\omega_{n},\mathbf{k})\gamma_{y}
G_{0}(i\omega_{n},\mathbf{k})\right], \label{Eq:SusceptZDef}
\end{eqnarray}
where $\gamma_{x}$, $\gamma_{y}$, and $\gamma_{z}$ are given by the
Eqs.~(\ref{Eq:gammaxDef})-(\ref{Eq:gammazDef}). Substituting the
fermion propagator and $\gamma_{i}$ into
Eqs.~(\ref{Eq:SusceptBotDef}) and (\ref{Eq:SusceptZDef}), and
performing integration over the azimuthal angle within the $x$-$y$
plane, we obtain
\begin{eqnarray}
\chi_{D}^{\bot} &=&N A^{2}B^{2}e^{2}T\frac{36}{\pi^{2}}\int dk_{\bot}
d|k_{z}|k_{\bot}^{5}k_{z}^{2}\nonumber
\\
&&\times\left(S_{A} - 4B^{2}k_{\bot}^{6}A^{2}
k_{z}^{4}S_{B}\right),\label{Eq:SusceptBotMid} \\
\chi_{D}^{z} &=& NB^{4}e^{2}T\frac{81}{\pi^{2}}\int
dk_{\bot}d|k_{z}|k_{\bot}^{9}\nonumber
\\
&&\times \left(S_{A} -
B^{4}k_{\bot}^{12}S_{B}\right).\label{Eq:SusceptZMid}
\end{eqnarray}
where
\begin{eqnarray}
S_{A} &=& \sum_{\omega_{n}} \frac{1}{\left(\omega_{n}^{2} +
E_{\mathbf{k}}^{2}\right)^{2}}, \\
S_{B} &=& \sum_{\omega_{n}} \frac{1}{\left(\omega_{n}^{2} +
E_{\mathbf{k}}^{2}\right)^{4}}.
\end{eqnarray}
After carrying out frequency summation, $S_{A}$ and $S_{B}$ can be
further written as
\begin{eqnarray}
S_{A} &=& \frac{1}{(2\pi T)^{4}}\frac{1}{2Y}
\left[\frac{\pi}{Y^{2}}\tanh\left(\pi Y\right) -
\frac{\pi^{2}}{Y}\frac{1}{\cosh^{2}\left(\pi Y\right)}\right],
\label{Eq:SAResult} \\
S_{B} &=& \frac{1}{(2\pi T)^{8}}\frac{1}{48Y^{3}}\left[
\frac{15\pi}{Y^{4}}\tanh\left(\pi Y\right) -
\frac{15\pi^{2}}{Y^{3}}\frac{1}{\cosh^{2}\left(\pi Y\right)}\right.\nonumber
\\
&&-\frac{12\pi^{3}}{Y^{2}}\frac{\tanh\left(\pi Y\right)}{\cosh^{2}
\left(\pi Y\right)} + \frac{2\pi^{4}}{Y}\frac{1}{\cosh^{4}\left(\pi
Y\right)} \nonumber
\\
&&\left.- \frac{4\pi^{4}}{Y}\frac{\tanh^{2}\left(\pi
Y\right)}{\cosh^{2}\left(\pi Y\right)}\right],
\label{Eq:SBResult}
\end{eqnarray}
where $Y=\frac{E_{\mathbf{k}}}{2\pi T}$. Substituting
Eqs.~(\ref{Eq:SAResult}) and (\ref{Eq:SBResult}) into
Eqs.~(\ref{Eq:SusceptBotMid}) and (\ref{Eq:SusceptZMid}), and then
invoking transformations
Eqs.~(\ref{Eq:TransformationsA})-(\ref{Eq:TransformationsC}),
followed by an integration over $\xi$, we find that the following
diamagnetic susceptibilities
\begin{eqnarray}
\chi_{D}^{\bot} &=& Nc_{\chi^{\bot}}\sqrt{A}e^{2}\sqrt{T}, \\
\chi_{D}^{z} &=& Nc_{\chi_{z}}\frac{B^{2/3}}{\sqrt{A}}e^{2}
T^{5/6},
\end{eqnarray}
where
\begin{eqnarray}
c_{\chi^{\bot}} &=& \frac{2}{77\sqrt{2}\pi^{2}}\int_{0}^{+\infty} dx
\left[\frac{47}{x^{1/2}}\tanh(x) - 47
\frac{x^{1/2}}{\cosh^{2}(x)} \right.\nonumber
\\
&& +24x^{3/2}\frac{\tanh(x)}{\cosh^{2}(x)}-4x^{5/2}
\frac{1}{\cosh^{4}(x)}\nonumber \\
&&\left.+ 8x^{5/2}
\frac{\tanh^{2}(x)}{\cosh^{2}(x)}\right], \label{Eq:CChiBotOri}
\\
c_{\chi_{z}} &=& \frac{\Gamma\left(\frac{5}{4}\right)
\Gamma\left(\frac{5}{3}\right)}{322\pi^{2}2^{1/6}
\Gamma\left(\frac{23}{12}\right)} \int_{0}^{+\infty}dx
x^{11/6}\left[\frac{2187}{x^{2}}\tanh\left(x\right)\right.\nonumber
\\
&& -\frac{1}{x}\frac{2187}{\cosh^{2}\left(x\right)} +
1728\frac{\tanh\left(x\right)}{\cosh^{2}\left(x\right)} -
\frac{288x}{\cosh^{4}\left(x\right)}\nonumber
\\
&&\left.+576x\frac{\tanh^{2}\left(x\right)}{\cosh^{2}
\left(x\right)}\right]. \label{Eq:CChiZOri}
\end{eqnarray}
Here, a transformation $E = 2Tx$ has been used. Notice that the
integrals (over $x$) appearing in the first terms of the integrands
of Eqs.~(\ref{Eq:CChiBotOri}) and (\ref{Eq:CChiZOri}) are actually
divergent. To regularize the integral, we need to introduce a cutoff
$\frac{\Lambda}{2T}$. After doing so, we find that the low-$T$
diamagnetic susceptibilities can be written as
\begin{eqnarray}
\chi_{D}^{\bot} &=& \chi_{D0}^{\bot}+Nc_{6}\sqrt{A}e^{2}\sqrt{T},
\\
\chi_{D}^{z} &=& \chi_{D0}^{z} + Nc_{7}
\frac{B^{2/3}}{\sqrt{A}}e^{2}T^{5/6},
\end{eqnarray}
where
\begin{eqnarray}
\chi_{D0}^{\bot} &=& N\frac{94}{77\pi^{2}}
\sqrt{A}e^{2}\sqrt{\Lambda}, \\
\chi_{D0}^{z} &=& N\frac{6561\Gamma\left(\frac{5}{4}\right)\Gamma
\left(\frac{5}{3}\right)}{1610\pi^{2}\Gamma\left(\frac{23}{12}\right)}
\frac{B^{2/3}}{\sqrt{A}}e^{2}\Lambda^{5/6},
\end{eqnarray}
and
\begin{widetext}
\begin{eqnarray}
c_{6}&=&-\frac{188}{77\sqrt{2}\pi^{2}} + \frac{94}{77\sqrt{2}
\pi^{2}}\int_{1}^{+\infty} dx\left[\frac{1}{x^{1/2}}
\tanh\left(x\right)-\frac{1}{x^{1/2}}\right]+\frac{94}{77
\sqrt{2}\pi^{2}}\int_{0}^{1} dx\frac{1}{x^{1/2}}
\tanh\left(x\right) \nonumber \\
&&+\frac{2}{77\sqrt{2}\pi^{2}}\int_{0}^{+\infty} dx\left[-
47x^{1/2}\frac{1}{\cosh^{2}\left(x\right)} +
24x^{3/2}\frac{\tanh\left(x\right)}{\cosh^{2}\left(x\right)}
- 4x^{5/2}\frac{1}{\cosh^{4}\left(x\right)}+8x^{5/2}
\frac{\tanh^{2}\left(x\right)}{\cosh^{2}\left(x\right)}\right]
\nonumber \\
&\approx& 0.1869, \\
c_{7} &=& -\frac{13122\Gamma\left(\frac{5}{4}\right)
\Gamma\left(\frac{5}{3}\right)}{1610\pi^{2}2^{1/6}
\Gamma\left(\frac{23}{12}\right)} +\frac{2187\Gamma
\left(\frac{5}{4}\right)\Gamma\left(\frac{5}{3}\right)}{322\pi^{2}
2^{1/6}\Gamma\left(\frac{23}{12}\right)}\int_{1}^{+\infty}
dx \left[\frac{1}{x^{1/6}}\tanh\left(x\right) -
\frac{1}{x^{1/6}}\right] + \frac{2187
\Gamma\left(\frac{5}{4}\right)
\Gamma\left(\frac{5}{3}\right)}{322\pi^{2}2^{1/6}
\Gamma\left(\frac{23}{12}\right)}\int_{0}^{1} dx
\frac{1}{x^{1/6}}\tanh\left( x\right)\nonumber \\
&&+\frac{\Gamma\left(\frac{5}{4}\right)\Gamma\left(\frac{5}{3}\right)}{322
\pi^{2}2^{1/6}\Gamma\left(\frac{23}{12}\right)}\int_{0}^{+\infty}
dx x^{11/6}\left[-\frac{2187}{x}\frac{1}{\cosh^{2}
\left(x\right)}+1728\frac{\tanh\left(x\right)}{\cosh^{2}
\left(x\right)}-288x\frac{1}{\cosh^{4}\left(x\right)} + 576x
\frac{\tanh^{2}\left( x\right)}{\cosh^{2}\left(x\right)}\right]
\nonumber \\
&\approx& -0.4131.
\end{eqnarray}
\end{widetext}

\section{Observable quantities renormalized by Coulomb interaction
\label{App:ObserQuanInteract}}

Here, we re-calculate the observable quantities by including the
corrections caused by the Coulomb interaction. The strategy is to
replace the constant model parameters appearing in these quantities
by the renormalized parameters, which are obtained from the
solutions of coupled RG equations.

\subsection{DOS}

The DOS of free fermions is
\begin{eqnarray}
\rho &\sim& \frac{1}{\sqrt{A} B^{2/3}} \omega^{1/6}.
\end{eqnarray}
Since parameters $A$ and $B$ are singularly renormalized by the
Coulomb interaction, the low-energy behavior of $\rho(\omega)$ is
certainly modified. After including the interaction corrections, we
get
\begin{eqnarray}
\frac{d\ln(\rho)}{d\ln(\omega)} &\sim&\frac{1}{6}+\frac{\ln\left(\frac{1}{\sqrt{A} B^{2/3}}\right)}{d\ln(\omega)}.
\label{Eq:RhoVRGA}
\end{eqnarray}
We have got  the RG equations of $A$ and $B$:
\begin{eqnarray}
\frac{dA}{d\ell}&=&C_{3}A, \label{Eq:BDiffell}
\\
\frac{dB}{d\ell}&=&C_{2}B. \label{Eq:ADiffell}
\end{eqnarray}
Using the transformation $\omega = \omega_{0}e^{-\ell}$, we find
that $A$ and $B$ depend on energy $\omega$ as follows
\begin{eqnarray}
\frac{d\ln(A)}{d\ln(\omega)}=-C_{3}, \label{Eq:BDiffOmega}
\\
\frac{d\ln(B)}{d\ln(\omega)}=-C_{2}, \label{Eq:ADiffOmega}
\end{eqnarray}
Substituting Eqs.~(\ref{Eq:BDiffOmega}) and (\ref{Eq:ADiffOmega})
into Eq.~(\ref{Eq:RhoVRGA}), we have
\begin{eqnarray}
\frac{d\ln(\rho)}{d\ln(\omega)} &\sim&
\frac{1}{6}+\frac{1}{2}C_{3} + \frac{2}{3}C_{2} .
\end{eqnarray}
It can be further approximately written as
\begin{eqnarray}
\frac{d\ln(\rho)}{d\ln(\omega)} &\sim& \frac{1}{6} +
\frac{1}{2}C_{3}^{*} + \frac{2}{3}C_{2}^{*} ,
\end{eqnarray}
which has the following solution
\begin{eqnarray}
\rho(\omega)\sim \omega^{\frac{1}{6}+\frac{1}{2}C_{3}^{*} +
\frac{2}{3}C_{2}^{*}} \sim \omega^{1/6+\eta_{1}},
\end{eqnarray}
where
\begin{eqnarray}
\eta_{1}=\frac{2}{3}C_{2}^{*}+\frac{1}{2}C_{3}^{*}.
\end{eqnarray}

\subsection{Specific heat}

In the non-interacting limit, we have found the specific heat
\begin{eqnarray}
C_{v} &\sim& \frac{T^{7/6}}{\sqrt{A}B^{2/3}}.
\end{eqnarray}
Once the interaction corrections are taken into account, parameters
$A$ and $B$ become $T$-dependent at finite $T$. One can verify that
\begin{eqnarray}
\frac{dC_{v}}{dT} &=& \frac{7}{6}+\frac{\ln\left(\frac{1}{\sqrt{A}B^{2/3}}\right)}
{d\ln(T)}.
\label{Eq:CvVRGA}
\end{eqnarray}
Using the transformation $T = T_{0}e^{-\ell}$ where $T_{0}$ is an
initial value, we use Eqs.~(\ref{Eq:BDiffell}) and
(\ref{Eq:ADiffell}) to obtain
\begin{eqnarray}
\frac{d\ln(A)}{d\ln(T)} = -C_{3}, \label{Eq:BDiffTemperature} \\
\frac{d\ln(B)}{d\ln(T)} = -C_{2}. \label{Eq:ADiffTemperature}
\end{eqnarray}
Substituting Eqs.~(\ref{Eq:BDiffTemperature}) and
(\ref{Eq:ADiffTemperature}) into Eq.~(\ref{Eq:CvVRGA}), we get
\begin{eqnarray}
\frac{d\ln(C_{v})}{d\ln(T)} &\sim& \frac{7}{6} +
\frac{2}{3}C_{2} + \frac{1}{2}C_{3}\nonumber
\\
&\sim&\frac{7}{6} + \frac{2}{3}C_{2}^{*} +
\frac{1}{2}C_{3}^{*}, \nonumber
\end{eqnarray}
which leads to the renormalized specific heat
\begin{eqnarray}
C_{v}(T)\sim T^{\frac{7}{6}+\frac{2}{3}C_{2}^{*} +
\frac{1}{2}C_{3}^{*}}\sim T^{7/6+\eta_{1}}.
\end{eqnarray}

\subsection{Compressibility}

In the non-interacting limit, the compressibility is
\begin{eqnarray}
\kappa \sim \frac{1}{\sqrt{A}B^{2/3}}T^{1/6}.
\end{eqnarray}
After including interaction corrections, we obtain
\begin{eqnarray}
\frac{d\ln(\kappa)}{d\ln(T)} &\sim&\frac{1}{6}+\frac{d\ln\left(\frac{1}{\sqrt{A}B^{2/3}}\right)}
{d\ln(T)}.\label{Eq:CompressibilityVRGA}
\end{eqnarray}
Substituting Eqs.~(\ref{Eq:BDiffTemperature}) and
(\ref{Eq:ADiffTemperature}) into Eq.~(\ref{Eq:CompressibilityVRGA})
leads to
\begin{eqnarray}
\frac{d\ln(\kappa)}{d\ln(T)} &\sim& \frac{1}{6} +
\frac{2}{3}C_{2}+\frac{1}{2}C_{3} \nonumber
\\
&\sim& \frac{1}{6} +
\frac{2}{3}C_{2}^{*}+\frac{1}{2}C_{3}^{*} .
\end{eqnarray}
Now the renormalized compressibility can be written as
\begin{eqnarray}
\kappa(T)\sim T^{\frac{1}{6} + \frac{2}{3}C_{2}^{*} +
\frac{1}{2}C_{3}^{*}}\sim T^{1/6+\eta_{1}}.
\end{eqnarray}

\subsection{Conductivities}

In the free case, the conductivities within the $x$-$y$ plane and
along $z$-axis are given by
\begin{eqnarray}
\sigma_{\bot\bot} \sim \frac{e^{2}}{\sqrt{A}}
\omega^{1/2},\quad \mathrm{and} \quad \sigma_{zz}
\sim\frac{\sqrt{A}e^{2}}{B^{2/3}}\omega^{1/6}.
\end{eqnarray}
Incorporating the interaction corrections of $A$, $B$, and $e$, we
similarly have
\begin{eqnarray}
\frac{d\ln(\sigma_{\bot\bot})}{d\ln(\omega)} &\sim&\frac{1}{2}
+\frac{d\ln\left(\frac{e^{2}}{\sqrt{A}}\right)}{d\ln(\omega)},
\label{Eq:ConductBotVRGA}
\\
\frac{d\ln(\sigma_{zz})}{d\ln(\omega)} &\sim& \frac{1}{6}
+\frac{d\ln\left(\frac{\sqrt{A}e^{2}}{B^{2/3}}\right)}{d\ln(\omega)}. \label{Eq:ConductZVRGA}
\end{eqnarray}
From the definition $g=\frac{\sqrt{4\pi}e}{\epsilon}$ and the RG
equation
\begin{eqnarray}
\frac{dg}{d\ell} = \left(\frac{1}{4}-\frac{NC_{\bot}}{2}\right)g,
\end{eqnarray}
we get
\begin{eqnarray}
\frac{de}{d\ell} = \left(\frac{1}{4}-\frac{NC_{\bot}}{2}\right)e.
\label{Eq:eDiffell}
\end{eqnarray}
Once again we utilize the transformation $\omega =
\omega_{0}e^{-\ell}$ to obtain
\begin{eqnarray}
\frac{d\ln(e)}{d\ln(\omega)} = -\left(\frac{1}{4} -
\frac{NC_{\bot}}{2}\right). \label{Eq:eDiffOmega}
\end{eqnarray}
Substituting Eqs.~(\ref{Eq:BDiffOmega}), (\ref{Eq:ADiffOmega}), and
(\ref{Eq:eDiffOmega}) into Eqs.~(\ref{Eq:ConductBotVRGA}) and
(\ref{Eq:ConductZVRGA}), we find that the conductivities satisfy the
equations as follows
\begin{eqnarray}
\frac{d\ln(\sigma_{\bot\bot})}{d\ln(\omega)} &\sim& \frac{1}{2} -
\frac{1}{2}+NC_{\bot} + \frac{1}{2}C_{3}\nonumber
\\
&\sim& \frac{1}{2}-\frac{1}{2}+NC_{\bot}^{*} +
\frac{1}{2}C_{3}^{*},
\\
\frac{d\ln(\sigma_{zz})}{d\ln(\omega)} &\sim& \frac{1}{6} -
\frac{1}{2}+NC_{\bot} +\frac{2}{3}C_{2}-\frac{1}{2}C_{3}\nonumber
\\
&\sim& \frac{1}{6} - \frac{1}{2} +
NC_{\bot}^{*} +\frac{2}{3}C_{2}^{*}-\frac{1}{2}C_{3}^{*}.
\end{eqnarray}
Solving these two equations, we now obtain the renormalized
conductivities
\begin{eqnarray}
\sigma_{\bot\bot}(\omega) &\sim& \omega^{\frac{1}{2} -
\frac{1}{2}+NC_{\bot}^{*} + \frac{1}{2}C_{3}^{*}}\sim
\omega^{1/2+\eta_{2}}, \\
\sigma_{zz}(\omega) &\sim& \omega^{\frac{1}{6}-\frac{1}{2} +
NC_{\bot}^{*} + \frac{2}{3}C_{2}^{*}-\frac{1}{2}C_{3}^{*}}\sim
\omega^{1/6+\eta_{3}},
\end{eqnarray}
where
\begin{eqnarray}
\eta_{2} &=& -\frac{1}{2}+NC_{\bot}^{*}+\frac{1}{2}C_{3}^{*}=0,
\\
\eta_{3} &=& -\frac{1}{2}+NC_{\bot}^{*} + \frac{2}{3}C_{2}^{*} -
\frac{1}{2}C_{3}^{*}.
\end{eqnarray}
We see that $\sigma_{\bot\bot}(\omega)$ remains qualitatively intact
after including the interaction corrections, whereas
$\sigma_{zz}(\omega)$ is considerably altered.

\begin{table}[htbp]
\caption{Dependence of $\eta_{1}$, $\eta_{2}$, $\eta_{3}$,
$\eta_{4}$, and $\eta_{5}$ on fermion flavor $N$.
\label{Table:EetaiFermionFlavor}}
\begin{center}
\begin{tabular}{|c|c|c|c|c|c|}
\hline\hline    $N$ & $\eta_{1}$ & $\eta_{2}$ & $\eta_{3}$ &
$\eta_{4}$ & $\eta_{5}$
\\
\hline          1& 0.07938 & 0  &  0.1057 & 0.01754 & -0.08815
\\
\hline          2& 0.04132 & 0  &  0.05518 & 0.009236 & -0.04594
\\
\hline          3& 0.02792 & 0  &  0.03731 & 0.006258 & -0.03105
\\
\hline          4& 0.02108 & 0  &  0.02818 & 0.004731 & -0.02345
\\
\hline          5& 0.01693 & 0  &  0.02264 & 0.003802 & -0.01883
\\
\hline          6& 0.01415 & 0  &  0.01892 &  0.003178 & -0.01574
\\
\hline          7& 0.01215  & 0 &  0.01625  & 0.00273 & -0.01352
\\
\hline          8& 0.01065  & 0 &  0.01424 & 0.002393 & -0.01184
\\
\hline          9& 0.009474  & 0 & 0.01267 & 0.00213 & -0.01054
\\
\hline          10& 0.008534  &  0& 0.01141 & 0.001919 & -0.009494
\\
\hline \hline
\end{tabular}
\end{center}
\end{table}

\subsection{Diamagnetic Susceptibilities}

In the free fermion system, the diamagnetic susceptibilities within
the $x$-$y$ plane and along $z$-axis satisfy
\begin{eqnarray}
\Delta\chi_{D}^{\bot} \sim \sqrt{A}e^{2}\sqrt{T} \quad
\mathrm{and}\quad \Delta\chi_{D}^{z} \sim
\frac{B^{2/3}}{\sqrt{A}} e^{2}T^{5/6}.
\end{eqnarray}
Here, $\Delta\chi_{D}^{\bot}=\chi_{D}^{\bot}-\chi_{D0}^{\bot}$ and
$\Delta\chi_{D}^{z} = \chi_{D}^{z}-\chi_{D0}^{z}$, where
$\chi_{D0}^{\bot}$ and $\chi_{D0}^{z}$ are the residual constant
value at $T=0$. Considering the influence of Coulomb interaction,
$\Delta\chi_{D}^{\bot}$ and $\Delta\chi_{D}^{z}$ become
\begin{eqnarray}
\frac{d\ln(\Delta\chi_{D}^{\bot})}{d\ln(T)} &\sim&
\frac{1}{2}+\frac{d\ln\left(\sqrt{A}e^{2}\right)}{d\ln(T)}, \label{Eq:SuspBotVRGA}
\\
\frac{d\ln(\Delta\chi_{D}^{z})}{d\ln(T)} &\sim& \frac{5}{6}
+\frac{d\ln\left(\frac{B^{2/3}}{\sqrt{A}} e^{2}\right)}{d\ln(T)}. \label{Eq:SuspZVRGA}
\end{eqnarray}
Employing the transformation $T = T_{0}e^{-\ell}$, we get
\begin{eqnarray}
\frac{d\ln(e)}{d\ln(T)} &=& -\left(\frac{1}{4}-\frac{NC_{\bot}}{2}\right).
\label{Eq:eDiffTemperature}
\end{eqnarray}
Substituting Eqs.~(\ref{Eq:BDiffTemperature}),
(\ref{Eq:ADiffTemperature}), and (\ref{Eq:eDiffTemperature}) into
Eqs.~(\ref{Eq:SuspBotVRGA}) and (\ref{Eq:SuspZVRGA}), it is easy to
find that
\begin{eqnarray}
\frac{d\ln(\Delta\chi_{D}^{\bot})}{d\ln(T)} &\sim& \frac{1}{2} -
\frac{1}{2}+NC_{\bot} - \frac{1}{2}C_{3}\nonumber
\\
&\sim&\frac{1}{2} -
\frac{1}{2}+NC_{\bot}^{*} - \frac{1}{2}C_{3}^{*},
\\
\frac{d\ln(\Delta\chi_{D}^{z})}{d\ln(T)} &\sim& \frac{5}{6} -
\frac{1}{2}+NC_{\bot} - \frac{2}{3}C_{2}+\frac{1}{2}C_{3}
\nonumber
\\
&\sim& \frac{5}{6} - \frac{1}{2} +
NC_{\bot}^{*} - \frac{2}{3}C_{2}^{*}+\frac{1}{2}C_{3}^{*},
\end{eqnarray}
which directly give rise to the following renormalized dynamical
susceptibilities
\begin{eqnarray}
\Delta\chi_{D}^{\bot}(T) &\sim& T^{\frac{1}{2}-\frac{1}{2} +
NC_{\bot}^{*} - \frac{1}{2}C_{3}^{*}}\sim T^{1/2+\eta_{4}}, \\
\Delta\chi_{D}^{z}(T) &\sim& T^{\frac{5}{6}-\frac{1}{2}+NC_{\bot}^{*} -
\frac{2}{3}C_{2}^{*} + \frac{1}{2}C_{3}^{*}}\sim T^{5/6+\eta_{5}},
\end{eqnarray}
where
\begin{eqnarray}
\eta_{4} &=& -\frac{1}{2}+NC_{\bot}^{*}-\frac{1}{2}C_{3}^{*},
\\
\eta_{5}&=&-\frac{1}{2}+NC_{\bot}^{*}-\frac{2}{3}C_{2}^{*} +
\frac{1}{2}C_{3}^{*}.
\end{eqnarray}

The values of $\eta_{1}$, $\eta_{2}$, $\eta_{3}$, $\eta_{4}$, and
$\eta_{5}$ for a series of flavors are given in
Table~\ref{Table:EetaiFermionFlavor}.

\section{Interplay between long-range and short-range interactions \label{App:InterplayCoulombFourFermion}}

Here, we study the interplay between the Coulomb interaction and
some types of four-fermion interactions, and analyze the nontrivial
physical effects of such an interplay.

\subsection{Fierz identity}

The four-fermion-type short-range interactions between fermions can
be generically written as
\begin{eqnarray}
S_{\psi^{4}} &=& \frac{1}{N}\int d\tau d^3\mathbf{x}
\left[\lambda_{0}\left(\psi^{\dag}\sigma_{0}\psi\right)^{2} +
\lambda_{1}\left(\psi^{\dag}\sigma_{1}\psi\right)^{2}\right.\nonumber
\\
&&\left.+\lambda_{2}\left(\psi^{\dag}\sigma_{2}\psi\right)^{2} +
\lambda_{3}\left(\psi^{\dag}\sigma_{3}\psi\right)^{2}\right],
\end{eqnarray}
where $\sigma_{0}$ is the identity matrix. As dictated by the Fierz
identity, the four coupling terms
$\left(\psi^{\dag}\sigma_{0}\psi\right)^{2} \equiv
\left(\psi^{\dag}\psi\right)^{2}$,
$\left(\psi^{\dag}\sigma_{1}\psi\right)^{2}$,
$\left(\psi^{\dag}\sigma_{2}\psi\right)^{2}$,
$\left(\psi^{\dag}\sigma_{3}\psi\right)^{2}$ are not independent
\cite{Roy17, RoyFoster18}. According to the Fierz identity, one can
find that \cite{Roy17, RoyFoster18}
\begin{widetext}
\begin{eqnarray}
\left[\psi^{\dag}(x)\sigma_{a}\psi(x)\right]
\left[\psi^{\dag}(y)\sigma_{b}\psi(y)\right] =
-\frac{1}{4}\mathrm{Tr}\left[\sigma_{a}\sigma_{c}\sigma_{b}
\sigma_{d}\right]\left[\psi^{\dag}(x) \sigma_{c}\psi(y)\right]
\left[\psi^{\dag}(y)\sigma_{d}\psi(x)\right],
\end{eqnarray}
where $a,b,c,d = 0,1,2,3$. Locality requires that $x = y$, which
leads to
\begin{eqnarray}
\left[\psi^{\dag}(x)\sigma_{a}\psi(x)\right]
\left[\psi^{\dag}(x)\sigma_{b}\psi(x)\right] =
-\frac{1}{4}\mathrm{Tr}\left[\sigma_{a}\sigma_{c}\sigma_{b}
\sigma_{d}\right]\left[\psi^{\dag}(x)\sigma_{c}\psi(x)\right]
\left[\psi^{\dag}(x)\sigma_{d}\psi(x)\right].
\end{eqnarray}
From this equation, we get
\begin{eqnarray}
3\left[\psi^{\dag}(x)\sigma_{0}\psi(x)\right]^{2}
+\left[\psi^{\dag}(x)\sigma_{1}\psi(x)\right]^{2}
+\left[\psi^{\dag}(x)\sigma_{2}\psi(x)\right]^{2}
+\left[\psi^{\dag}(x)\sigma_{3}\psi(x)\right]^{2}&=&0,
\label{Eq:FourFermionRelation1}
\\
\left[\psi^{\dag}(x)\sigma_{0}\psi(x)\right]^{2}
+3\left[\psi^{\dag}(x)\sigma_{1}\psi(x)\right]^{2}
-\left[\psi^{\dag}(x)\sigma_{2}\psi(x)\right]^{2}
-\left[\psi^{\dag}(x)\sigma_{3}\psi(x)\right]^{2}&=&0,
\label{Eq:FourFermionRelation2}
\\
\left[\psi^{\dag}(x)\sigma_{0}\psi(x)\right]^{2}
-\left[\psi^{\dag}(x)\sigma_{1}\psi(x)\right]^{2}
+3\left[\psi^{\dag}(x)\sigma_{2}\psi(x)\right]^{2}
-\left[\psi^{\dag}(x)\sigma_{3}\psi(x)\right]^{2}&=&0,
\label{Eq:FourFermionRelation3}
\\
\left[\psi^{\dag}(x)\sigma_{0}\psi(x)\right]^{2}
-\left[\psi^{\dag}(x)\sigma_{1}\psi(x)\right]^{2}
-\left[\psi^{\dag}(x)\sigma_{2}\psi(x)\right]^{2} +
3\left[\psi^{\dag}(x)\sigma_{3}\psi(x)\right]^{2}&=&0.
\label{Eq:FourFermionRelation4}
\end{eqnarray}
\end{widetext}
Defining
\begin{eqnarray}
X = \left(\begin{array}{c}
\left(\psi^{\dag}\sigma_{0}\psi\right)^{2}
\\
\left(\psi^{\dag}\sigma_{1}\psi\right)^{2}
\\
\left(\psi^{\dag}\sigma_{2}\psi\right)^{2}
\\
\left(\psi^{\dag}\sigma_{3}\psi\right)^{2}
\end{array}\right)
\end{eqnarray}
and
\begin{eqnarray}
F = \left(\begin{array}{cccc} 3 & 1 & 1 & 1
\\
1 & 3 & -1 & -1
\\
1 & -1 & 3 & -1
\\
1 & -1 & 3 & -1
\end{array}\right),
\end{eqnarray}
we now can re-write
Eqs.~(\ref{Eq:FourFermionRelation1})-(\ref{Eq:FourFermionRelation4})
in a more compact form
\begin{eqnarray}
FX=0.
\end{eqnarray}
Solving this equation gives rise to
\begin{eqnarray}
-\left(\psi^{\dag}\sigma_{0}\psi\right)^{2} =
\left(\psi^{\dag}\sigma_{1}\psi\right)^{2} =
\left(\psi^{\dag}\sigma_{2}\psi\right)^{2} =
\left(\psi^{\dag}\sigma_{3}\psi\right)^{2}. \nonumber \\
\label{Eq:FierzIdentityRS}
\end{eqnarray}
Similar to previous works on four-fermion interactions \cite{Roy17,
RoyFoster18, Roy16}, we will use Eq.~(\ref{Eq:FierzIdentityRS}) to
carry out RG calculations.

\subsection{Physical meaning of fermion bilinears}

Upon decoupling the four-fermion interaction terms and then taking
the expectation value, we obtain four different fermion bilinears:
\begin{eqnarray}
&&\Delta_{0}\equiv \left<\psi^{\dag}\sigma_{0}\psi\right> \equiv
\left<\psi^{\dag}\psi \right>, \qquad
\Delta_{1}\equiv\left<\psi^{\dag}\sigma_{1}\psi\right>,\nonumber
\\
&&\Delta_{2}\equiv\left<\psi^{\dag}\sigma_{2}\psi\right>, \qquad
\Delta_{3}\equiv\left<\psi^{\dag}\sigma_{3}\psi\right>.
\end{eqnarray}
When the coupling parameter $\lambda_{i}$, where $i=0,1,2,3$,
diverges at low energies, the corresponding $\Delta_{i}$ acquires a
finite value. Each fermion billinear has its special physical
meaning. At the TQCP, the original fermion dispersion is given by
\begin{eqnarray}
E = \pm\sqrt{B^{2}d_{1}^{2}(\mathbf{k}) + B^{2}
d_{2}^{2}(\mathbf{k})+A^{2}d_{3}^{2}(\mathbf{k})},
\end{eqnarray}
which vanishes at the point $(0,0,0)$. This dispersion will be
altered if $\Delta_{i}$ becomes finite.

If $\Delta_{0}\equiv\left<\psi^{\dag}\sigma_{0}\psi\right>$ acquires
a finite value, the fermion dispersion becomes
\begin{eqnarray}
E = \Delta_{0}\pm \sqrt{B^{2}d_{1}^{2}(\mathbf{k}) +
B^{2}d_{2}^{2}(\mathbf{k})+A^{2}d_{3}^{2}(\mathbf{k})},
\end{eqnarray}
which vanishes not at $(0,0,0)$ but on the surface $\sqrt{B^{2}
d_{1}^{2}(\mathbf{k}) + B^{2} d_{2}^{2}(\mathbf{k}) +
A^{2}d_{3}^{2}(\mathbf{k})} = |\Delta_{0}|$. It is easy to see that
$\Delta_{0}\equiv \left<\psi^{\dag} \sigma_{0} \psi\right>$
represents finite chemical potential.

If $\Delta_{1} \equiv \left<\psi^{\dag}\sigma_{1}\psi\right>$
becomes finite, the fermion dispersion is changed into
\begin{eqnarray}
E = \pm\sqrt{\left(Bd_{1}(\mathbf{k}) + \Delta_{1}\right)^{2} +
B^{2}d_{2}^{2}(\mathbf{k})+A^{2}d_{3}^{2}(\mathbf{k})}.
\end{eqnarray}
This dispersion vanishes at three different points:
\begin{eqnarray}
&&\left\{\begin{array}{l}
k_{ax}=-\left(\frac{\Delta_{1}}{B}\right)^{1/3}
\\
k_{ay}=0
\\
k_{az}=0
\end{array}\right.,\qquad
\left\{\begin{array}{l}
k_{bx}=\frac{1}{2}\left(\frac{\Delta_{1}}{B}\right)^{1/3}
\\
k_{by}=\frac{\sqrt{3}}{2}\left(\frac{\Delta_{1}}{B}\right)^{1/3}
\\
k_{bz}=0
\end{array}\right.,\nonumber
\\
&&\left\{\begin{array}{l}
k_{cx}=\frac{1}{2}\left(\frac{\Delta_{1}}{B}\right)^{1/3}
\\
k_{cy}=-\frac{\sqrt{3}}{2}\left(\frac{\Delta_{1}}{B}\right)^{1/3}
\\
k_{cz}=0
\end{array}\right..
\end{eqnarray}
One can expand the dispersion in the vicinity of these three points.
We define $\mathbf{K}$ as the momentum relative to the gapless
points, and then find that
\begin{eqnarray}
E = \pm\sqrt{9B^{2/3}\Delta_{1}^{4/3}K_{\bot}^{2} +
A^{2}K_{z}^{4}},
\end{eqnarray}
which is linear along two directions and quadratic along the third
one. Notice that the above effective dispersion is the same for all
these three gapless points. One can identify $\Delta_{1}$ as a
nematic order parameter, because the positions of these three points
break the equivalence between $x$- and $y$-axis.

If $\Delta_{2}=\left<\psi^{\dag}\sigma_{2}\psi\right>$ becomes
finite, the fermion dispersion has the new form
\begin{eqnarray}
E = \pm\sqrt{Bd_{1}^{2}(\mathbf{k}) + \left(Bd_{2}(\mathbf{k}) +
\Delta_{2}\right)^{2}+A^{2}d_{3}^{2}(\mathbf{k})}.
\end{eqnarray}
This energy also vanishes at three gapless points
\begin{eqnarray}
&&\left\{\begin{array}{l}
k_{ax}=0
\\
k_{ay}=-\left(\frac{\Delta_{2}}{B}\right)^{1/3}
\\
k_{az}=0
\end{array}\right.,\qquad
\left\{\begin{array}{l} k_{bx} = \frac{\sqrt{3}}{2}
\left(\frac{\Delta_{2}}{B}\right)^{1/3}
\\
k_{by}=\frac{1}{2}\left(\frac{\Delta_{2}}{B}\right)^{1/3}
\\
k_{bz}=0
\end{array}\right.,\qquad\nonumber
\\
&&\left\{\begin{array}{l}
k_{cx}=-\frac{\sqrt{3}}{2}\left(\frac{\Delta_{2}}{B}
\right)^{1/3}
\\
k_{cy}=\frac{1}{2}\left(\frac{\Delta_{2}}{B} \right)^{1/3}
\\
k_{cz}=0
\end{array}\right..
\end{eqnarray}
Introducing $\mathbf{K}$ as the momentum relative to gapless points,
the effective dispersion becomes
\begin{eqnarray}
E \pm\sqrt{9B^{2/3} \Delta_{2}^{4/3}K_{\bot}^{2} +
A^{2}K_{z}^{4}},
\end{eqnarray}
which is also linear along two directions and quadratic along the
third. Apparently, the impact of finite $\Delta_2$ is qualitatively
very similar to that of finite $\Delta_1$. Thus, $\Delta_{2}$ can
also be identified as a nematic order parameter.

If $\Delta_{3}=\left<\psi^{\dag}\sigma_{3}\psi\right>$ becomes
finite. the fermion dispersion reads
\begin{eqnarray}
E = \pm\sqrt{B^{2}d_{1}^{2}(\mathbf{k}) +
B^{2}d_{2}^{2}(\mathbf{k})+\left(Ad_{3}(\mathbf{k}) +
\Delta_{3}\right)^{2}}.
\end{eqnarray}
The energy gap for band insultor vanishes continuously upon
approaching the TQCP when $\Delta_3 = 0$. In contrast, this gap
vanishes abruptly at the TQCP once $\Delta_3 \neq 0$. Therefore, the
TQPT between triple-WSM and BI becomes first order if
$\left<\psi^{\dag} \sigma_{3}\psi\right> \neq 0$. Recently, Roy
\emph{et al.} \cite{Roy17} have studied the four-fermion interaction
$\left(\psi^{\dag}\sigma_{3}\psi\right)^{2}$ at the TQCP between
triple-WSM and BI. This interaction is found to be irrelevant if the
initial strength is small. If the initial strength is larger than
some critical value, the mean value
$\left<\psi^{\dag}\sigma_{3}\psi\right>$ becomes finite and the
original continuous TQPT becomes first order \cite{Roy17}.

Below we will combine the Coulomb interaction with every possible
four-fermion interaction. The interplay between Coulomb interaction
and $\left(\psi^{\dag}\sigma_{1}\psi\right)^{2}$ leads to
qualitatively the same low-energy properties as that between Coulomb
interaction and $\left(\psi^{\dag}\sigma_{2}\psi\right)^{2}$.
Therefore, we only need to consider the former case.

\begin{figure*}[htbp]
\center
\includegraphics[width=5in]{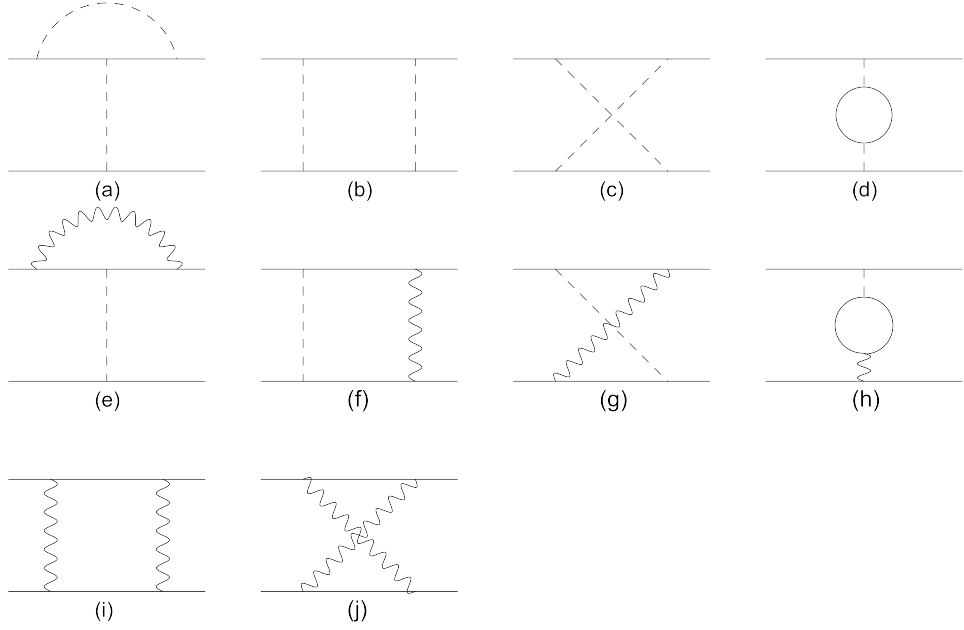}
\caption{Feynman diagrams for the vertex corrections to four-fermion
interaction. Solid, wavy, and dashed lines represent the fermion
propagator, Coulomb interaction, and four-fermion interaction,
respectively. \label{Fig:VertexCorrection}}
\end{figure*}

\subsection{Interplay with $(\psi^{\dag}\sigma_{1}\psi)^{2}$}

We will derive the flow equation for the coupling parameter
$\lambda_1$. For this purpose, we need to compute the Feynman
diagrams for the vertex corrections to
$(\psi^{\dag}\sigma_{1}\psi)^{2}$. All the involved diagrams are
presented in Fig.~\ref{Fig:VertexCorrection}.

The contribution from the diagram of
Fig.~\ref{Fig:VertexCorrection}(a) is given by
\begin{eqnarray}
\delta \lambda_{1}^{(1)}\sigma_{1} &=& \frac{4\lambda_{1}^{2}}{N}\int'
\frac{d\omega}{2\pi}\frac{d^3\mathbf{k}}{(2\pi)^{3}}
\sigma_{1}G_{0}\left(i\omega,\mathbf{k}\right)\sigma_{1}
G_{0}\left(i\omega,\mathbf{k}\right)\nonumber
\\
&&\times\sigma_{1}.
\label{Eq:VertexCorrectionNem1}
\end{eqnarray}
Substituting Eq.~(\ref{Eq:FermionPropagator}) into
Eq.~(\ref{Eq:VertexCorrectionNem1}) and then employing the
transformations
Eqs.~(\ref{Eq:TransformationsA})-(\ref{Eq:TransformationsC}), we
obtain
\begin{eqnarray}
\delta \lambda_{1}^{(1)} = -\frac{\lambda_{1}^{2}}{N}
\frac{5\Gamma\left(\frac{5}{4}\right)\Gamma\left(\frac{4}{3}\right)
\Lambda^{1/6}}{12\pi^{2}\Gamma\left(\frac{19}{12}\right)
\sqrt{A}B^{2/3}}\ell.
\end{eqnarray}
Diagrams shown in Figs.~\ref{Fig:VertexCorrection}(b) and
~\ref{Fig:VertexCorrection}(c) give rise to the correction
\begin{eqnarray}
V_{1}^{(2)+(3)} &=& \frac{4\lambda_{1}^{2}}{N}\int'\frac{d\omega}{2\pi}
\frac{d^3\mathbf{k}}{(2\pi)^{3}}\left(\psi^{\dag}\sigma_{1}
G_{0}(i\omega,\mathbf{k})\sigma_{1}\psi\right)\nonumber\\
&&\times\big\{\psi^{\dag}
\left[\sigma_{1}G_{0}(i\omega,\mathbf{k})\sigma_{1}\right.\nonumber
\\
&&\left. + \sigma_{1}
G_{0}(-i\omega,-\mathbf{k})\sigma_{1}\right]\psi\big\}\nonumber
\\
&=&\frac{\lambda_{1}^{2}}{N}\frac{\Gamma\left(\frac{1}{3}\right)
\Gamma\left(\frac{5}{4}\right)\Lambda^{1/6}}{12\pi^{2}
\Gamma\left(\frac{19}{12}\right)\sqrt{A}B^{2/3}}\ell
\left(\psi^{\dag}\sigma_{3}\psi\right)^{2}.
\end{eqnarray}
Making use of Fierz identity, we further have
\begin{eqnarray}
V_{1}^{(2)+(3)} = \frac{\lambda_{1}^{2}}{N}\frac{\Gamma\left(\frac{1}{3}
\right)\Gamma\left(\frac{5}{4}\right)\Lambda^{1/6}}{12\pi^{2}
\Gamma\left(\frac{19}{12}\right)\sqrt{A}B^{2/3}}\ell
\left(\psi^{\dag}\sigma_{1}\psi\right)^{2}.
\end{eqnarray}
Similarly, we get
\begin{eqnarray}
\delta \lambda_{1}^{(2)+(3)} = \frac{\lambda_{1}^{2}}{N}
\frac{\Gamma\left(\frac{1}{3}\right)\Gamma\left(\frac{5}{4}\right)
\Lambda^{1/6}}{12\pi^{2}\Gamma\left(\frac{19}{12}\right)
\sqrt{A}B^{2/3}}\ell.
\end{eqnarray}
The correction of diagram Fig.~\ref{Fig:VertexCorrection}(d) is given by
\begin{eqnarray}
\delta \lambda_{1}^{(4)} &=& -2N\frac{\lambda_{1}^{2}}{N}\int'
\frac{d\omega}{2\pi}\frac{d^{3}\mathbf{k}}{(2\pi)^{3}}
\mathrm{Tr}\left[\sigma_{1}G_{0}\left(i\omega,\mathbf{k}\right)
\sigma_{1}\right.\nonumber
\\
&&\left.\times G_{0}\left(i\omega,\mathbf{k}\right)\right]\nonumber
\\
&=&
\lambda_{1}^{2}\frac{5\Gamma\left(\frac{5}{4}\right)
\Gamma\left(\frac{4}{3}\right) \Lambda^{1/6}}{12\pi^{2}
\Gamma\left(\frac{19}{12}\right)\sqrt{A}B^{2/3}}\ell.
\end{eqnarray}
The diagram shown in Fig.~\ref{Fig:VertexCorrection}(e) yields
\begin{eqnarray}
\delta \lambda_{1}^{(5)} \sigma_{1} &=& -2g^{2}\lambda_{1}\int'
\frac{d\Omega}{2\pi}\frac{d^3\mathbf{q}}{(2\pi)^{3}}
G_{0}(i\Omega,\mathbf{q})\sigma_{1}G_{0}(i\Omega,\mathbf{q})\nonumber
\\
&&\times D_{0}(i\Omega,\mathbf{q}).
\end{eqnarray}
Carrying out direct calculations, we obtain
\begin{eqnarray}
\delta \lambda_{1}^{(5)} = \lambda_{1}C_{4}\ell,
\end{eqnarray}
where
\begin{eqnarray}
C_{4} &=& \frac{g^{2}}{24\pi^{2}\sqrt{A\Lambda}}
\int_{0}^{+\infty}d\xi\frac{2+\xi^{2}}{\xi^{1/3}
\left(1+\xi^{2}\right)^{13/12}}\nonumber
\\
&&\times\frac{1}{\xi^{2/3}
\left(1+\xi^{2}\right)^{1/6}+\zeta}.
\end{eqnarray}
For diagrams given by Figs.~\ref{Fig:VertexCorrection}(f) and \ref{Fig:VertexCorrection}(g), we get
\begin{eqnarray}
V_{1}^{(6)+(7)} &=& -4g^{2}\lambda_{1}\int'\frac{d\Omega}{2\pi}
\frac{d^3\mathbf{q}}{(2\pi)^{3}}\left(\psi_{\sigma}^{\dag}\sigma_{1}
G_{0}(i\Omega,\mathbf{q})\psi_{\sigma}\right)\nonumber
\\
&&\times\left\{\psi^{\dag}
\left[G_{0}(i\Omega,\mathbf{q})\sigma_{1} + \sigma_{1}
G_{0}(-i\Omega,-\mathbf{q})\right]\psi\right\}\nonumber
\\
&&\times D_{0}(i\Omega,\mathbf{q})\nonumber
\\
&\approx& \lambda_{1}C_{5}\ell\left(\psi^{\dag}\sigma_{3}
\psi\right)^{2},
\label{Eq:VertexCorrectionNem6and7}
\end{eqnarray}
where
\begin{eqnarray}
C_{5} &=& -\frac{g^{2}}{12\pi^{2}\sqrt{A\Lambda}}
\int_{0}^{+\infty}d\xi \frac{\xi^{5/3}}{\left(1 +
\xi^{2}\right)^{13/12}}\nonumber
\\
&&\times\frac{1}{\xi^{2/3}
\left(1+\xi^{2}\right)^{1/6} +\zeta}.
\end{eqnarray}
Fierz identity allows us to recast
Eq.~(\ref{Eq:VertexCorrectionNem6and7}) in the form
\begin{eqnarray}
V_{1}^{(6)+(7)} = \lambda_{1}C_{5}\ell\left(\psi^{\dag}\sigma_{1}
\psi\right)^{2},
\end{eqnarray}
which indicates that
\begin{eqnarray}
\delta\lambda_{1}^{(6)+(7)} = \lambda_{1}C_{5}\ell.
\end{eqnarray}
The correction due to the diagram shown in
Fig.~\ref{Fig:VertexCorrection}(h) is found to vanish, namely
\begin{eqnarray}
\delta\lambda_{1}^{(8)} &=& 2\lambda_{1}g^{2}\int'\frac{d\omega}{2\pi}
\frac{d^3\mathbf{k}}{(2\pi)^{3}}
\mathrm{Tr}\left[G_{0}(i\omega,\mathbf{k})\sigma_{1}\right.\nonumber
\\
&&\left.\times G_{0}(i\omega+i\Omega,\mathbf{k}+\mathbf{q})\right]
D_{0}(i\Omega,\mathbf{q})\nonumber
\\
&=&0.
\end{eqnarray}
Figs.~\ref{Fig:VertexCorrection}(i) and \ref{Fig:VertexCorrection}(j) induce the
contribution
\begin{eqnarray}
V_{1}^{(9)+(10)} &=& 4g^{4}\int'\frac{d\Omega}{2\pi}
\frac{d^3\mathbf{q}}{(2\pi)^{3}}\left(\psi^{\dag}
G_{0}\left(i\Omega,\mathbf{q}\right)\psi\right)D_{0}(i\Omega,\mathbf{q})\nonumber
\\
&&\times\left\{\psi^{\dag}\left[G_{0}\left(i\Omega,\mathbf{q}\right) +
G_{0}(-i\Omega,-\mathbf{q})\right]\psi\right\}\nonumber
\\
&&\times D_{0}(i\Omega,\mathbf{q})
\nonumber \\
&=&\frac{\sqrt{A}B^{2/3}}{\Lambda^{1/6}}C_{6}\ell
\left(\psi^{\dag}\sigma_{3}\psi\right)^{2}, \label{Eq:VertexCorrectionNem9and10}
\end{eqnarray}
where
\begin{eqnarray}
C_{6} &=& \frac{g^{4}}{6\pi^{2}A\Lambda} \int_{0}^{+\infty} d\xi
\frac{1}{\xi^{1/3}\left(1+\xi^{2}\right)^{7/12}}\nonumber
\\
&&\times\frac{1}{\left[\xi^{2/3}\left(1+\xi^{2}\right)^{1/6}
+ \zeta\right]^{2}}. \label{Eq:C6Expression}
\end{eqnarray}
We learn from Fierz identity that
Eq.~(\ref{Eq:VertexCorrectionNem9and10}) can also be written as
\begin{eqnarray}
V_{1}^{(9)+(10)} = \frac{\sqrt{A}B^{2/3}}{\Lambda^{1/6}}
C_{6}\ell\left(\psi^{\dag}\sigma_{1}\psi\right)^{2},
\end{eqnarray}
which gives rise to
\begin{eqnarray}
\delta\lambda_{1}^{(9)+(10)} = \frac{\sqrt{A}
B^{2/3}}{\Lambda^{1/6}}C_{6}\ell.
\end{eqnarray}

\begin{figure*}[htbp]
\center
\includegraphics[width=6.3in]{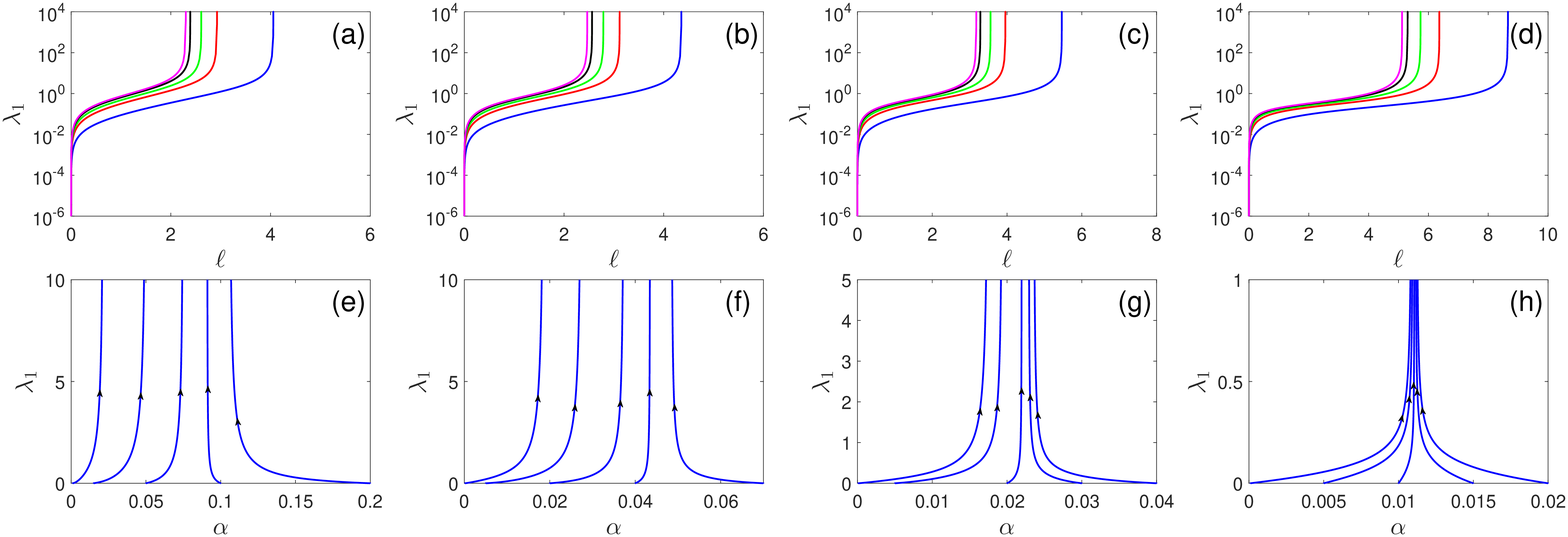}
\caption{(a)-(d): Dependence of $\lambda_{1}$ on $\ell$ at $N=1$,
$N=2$, $N=4$, $N=8$, respectively. Blue, red, green, black, magenta
lines correspond to $\alpha_{0}=0.01, 0.05, 0.1, 0.2, 0.3$,
respectively. (e)-(h): Flow diagrams on the $\alpha$-$\lambda_{1}$
plane at $N=1$, $N=2$, $N=4$, and $N=8$, respectively. Here, we take
$\beta_{0}=0.1$. \label{Fig:FlowDiagramFFlambda1}}
\end{figure*}

The total corrections to coupling parameter $\lambda_{1}$ are
\begin{eqnarray}
\delta\lambda_{1} &=& \delta\lambda_{1}^{(1)} + \delta
\lambda_{1}^{(2)+(3)}+\delta\lambda_{1}^{(4)} +\delta
\lambda_{1}^{(5)}+\delta\lambda_{1}^{(6)+(7)}\nonumber
\\
&&+\delta\lambda_{1}^{(8)}
+ \delta\lambda_{1}^{(9)+(10)}\nonumber
\\
&=&\left(1-\frac{2}{5N}\right)\lambda_{1}^{2}
\frac{5\Gamma\left(\frac{5}{4}\right)\Gamma\left(\frac{4}{3}\right)
\Lambda^{1/6}}{12\pi^{2}\Gamma\left(\frac{19}{12}\right)
\sqrt{A}B^{2/3}}\ell \nonumber
\\
&&+ \lambda_{1}\left(C_{4}+C_{5}\right)
\ell + \frac{\sqrt{A}B^{2/3}}{\Lambda^{1/6}}
C_{6}\ell. \label{Eq:TotalCorrectionlambda1}
\end{eqnarray}
Including these corrections to the original action leads to
\begin{eqnarray}
S_{\psi^{4}} &=& \frac{1}{N}\left(\lambda_{1} + \delta
\lambda_{1}\right)\int\frac{d\omega_{1}}{2\pi}\frac{d^{3}
\mathbf{k}_{1}}{(2\pi)^{3}}\frac{d\omega_{2}}{2\pi}
\frac{d^{3}\mathbf{k}_{2}}{(2\pi)^{3}}\frac{d\omega_{3}}{2\pi}
\frac{d^{3}\mathbf{k}_{3}}{(2\pi)^{3}}\nonumber
\\
&&\times\psi^{\dag}_{a}(\omega_{1},\mathbf{k}_{1})
\sigma_{1}\psi_{a}(\omega_{2},\mathbf{k}_{2})
\psi^{\dag}_{a}(\omega_{3},k_{3})\sigma_{1}\nonumber
\\
&& \times \psi_{a}(\omega_{1}-\omega_{2}+\omega_{3}, \mathbf{k}_{1}
- \mathbf{k}_{2}+\mathbf{k}_{3}).
\end{eqnarray}
Employing the transformations
Eqs.~(\ref{Eq:ScalingOmega})-(\ref{Eq:Scalingpsi}), we convert it
into
\begin{eqnarray}
S_{\psi'^{4}} &\approx& \frac{1}{N}\left(\lambda_{1}-\lambda_{1}
\frac{\ell}{6}+\delta\lambda_{1}\right)\int\frac{d\omega_{1}'}{2\pi}
\frac{d^{3}\mathbf{k}_{1}'}{(2\pi)^{3}}\frac{d\omega_{2}'}{2\pi}
\frac{d^{3}\mathbf{k}_{2}'}{(2\pi)^{3}}\nonumber
\\
&&\times\frac{d\omega_{3}'}{2\pi}
\frac{d^{3}\mathbf{k}_{3}'}{(2\pi)^{3}}
\psi'^{\dag}_{a}(\omega_{1}',\mathbf{k}_{1}')
\sigma_{1}\psi'_{a}(\omega_{2}',\mathbf{k}_{2}')
\psi'^{\dag}_{a}(\omega_{3}',k_{3}')  \nonumber
\\
&&\times\sigma_{1}\psi'_{a}(\omega_{1}'-\omega_{2}'+\omega_{3}',
\mathbf{k}_{1}'-\mathbf{k}_{2}'+\mathbf{k}_{3}').
\end{eqnarray}
By introducing a renormalized coupling parameter
\begin{eqnarray}
\lambda_{1}' = \lambda_{1}-\lambda_{1}
\frac{\ell}{6}+\delta\lambda_{1}, \label{Eq:Scalinglambda1}
\end{eqnarray}
the original form of the action can be recovered. From
Eqs.~(\ref{Eq:TotalCorrectionlambda1}) and
(\ref{Eq:Scalinglambda1}), we obtain the following flow equation for
$\lambda_{1}$:
\begin{eqnarray}
\frac{d\lambda_{1}}{d\ell} &=& -\frac{1}{6}\lambda_{1} +
\left(1-\frac{2}{5N}\right)\lambda_{1}^{2} + \left(C_{4}+C_{5} -
\frac{1}{2}C_{3}\right.\nonumber
\\
&&\left.-\frac{2}{3}C_{2}\right)\lambda_{1}+C_{7},
\end{eqnarray}
where
\begin{eqnarray}
C_{7} = \frac{5\Gamma\left(\frac{5}{4}\right)
\Gamma\left(\frac{4}{3}\right)}{12\pi^{2}
\Gamma\left(\frac{19}{12}\right)}C_{6}.
\end{eqnarray}
In the derivation of the above equation, we have made the
redefinition
\begin{eqnarray}
\frac{5\Gamma\left(\frac{5}{4}\right)\Gamma\left(\frac{4}{3}\right)
\Lambda^{1/6}}{12\pi^{2}\Gamma\left(\frac{19}{12}\right)
\sqrt{A}B^{2/3}}\lambda_{1} \rightarrow \lambda_{1},
\end{eqnarray}

In Figs.~\ref{Fig:FlowDiagramFFlambda1}(a)-\ref{Fig:FlowDiagramFFlambda1}(d), we show the RG flow
of $\lambda_{1}$ due to the interplay of four-fermion interaction
$\left(\psi^{\dag}\sigma_{1}\psi\right)^{2}$ and Coulomb
interaction. The corresponding flow diagrams on the
$\alpha$-$\lambda_{1}$ plane are displayed in
Figs.~\ref{Fig:FlowDiagramFFlambda1}(e)-\ref{Fig:FlowDiagramFFlambda1}(h). We observe that,
$\lambda_1$ increases quickly from zero as $\ell$ grows and goes to
infinity at a finite value $\ell_{c}$. The fermion bilinear
$\psi^{\dag}\sigma_{1}\psi$ acquires a finite expectation value,
i.e., $\langle\psi^{\dag}\sigma_{1}\psi\rangle \neq 0$. Now a
nematic order develops in the system, and splits the original
band-touching point into three different touching points. Around the
new points, the fermion dispersion is linear along two out of three
directions and quadratic along the rest one.

\subsection{Interplay with $(\psi^{\dag}\sigma_{3}\psi)^{2}$}

The correction from Fig.~\ref{Fig:VertexCorrection}(a) is
given by
\begin{eqnarray}
\delta \lambda_{3}^{(1)}\sigma_{3} &=& \frac{4\lambda_{3}^{2}}{N}
\int'\frac{d\omega}{2\pi}\frac{d^3\mathbf{k}}{(2\pi)^{3}}
\sigma_{3}G_{0}\left(i\omega,\mathbf{k}\right)\sigma_{3}\nonumber
\\
&&\times G_{0}\left(i\omega,\mathbf{k}\right)\sigma_{3}.
\label{Eq:VertexCorrectionTQPTPara1}
\end{eqnarray}
Substituting Eq.~(\ref{Eq:FermionPropagator}) into
Eq.~(\ref{Eq:VertexCorrectionTQPTPara1}) and carrying direct
calculations, we find
\begin{eqnarray}
\delta \lambda_{3}^{(1)} = -\frac{\lambda_{3}^{2}}{N} \frac{\Gamma
\left(\frac{5}{4}\right)\Gamma\left(\frac{4}{3}\right)
\Lambda^{1/6}}{3\pi^{2}\Gamma\left(\frac{19}{12}\right)
\sqrt{A}B^{2/3}}\ell.
\end{eqnarray}
The contribution induced by Figs.~\ref{Fig:VertexCorrection}(b)
and \ref{Fig:VertexCorrection}(c) can be written as
\begin{eqnarray}
V_{3}^{(2)+(3)} &=& \frac{4\lambda_{3}^{2}}{N}\int'\frac{d\omega}{2\pi}
\frac{d^3\mathbf{k}}{(2\pi)^{3}}\left(\psi^{\dag}\sigma_{3}
G_{0}(i\omega,\mathbf{k})\sigma_{3}\psi\right)\nonumber
\\
&&\times\left\{\psi^{\dag}
\left[\sigma_{3}G_{0}(i\omega,\mathbf{k})\sigma_{3}\right.\right.\nonumber
\\
&&\left.\left. + \sigma_{3}
G_{0}(-i\omega,-\mathbf{k})\sigma_{3}\right]\psi\right\}\nonumber
\\
&=&\frac{\lambda_{3}^{2}}{N}\frac{\Gamma\left(\frac{1}{3}\right)
\Gamma\left(\frac{5}{4}\right)\Lambda^{1/6}}{12\pi^{2}
\Gamma\left(\frac{19}{12}\right)\sqrt{A}B^{2/3}}\ell
\left(\psi^{\dag}\sigma_{3}\psi\right)^{2},
\end{eqnarray}
which represents
\begin{eqnarray}
\delta \lambda_{3}^{(2)+(3)} = \frac{\lambda_{3}^{2}}{N}
\frac{\Gamma\left(\frac{1}{3}\right)\Gamma\left(\frac{5}{4}\right)
\Lambda^{1/6}}{12\pi^{2}\Gamma\left(\frac{19}{12}\right)
\sqrt{A}B^{2/3}}\ell.
\end{eqnarray}
Fig.~\ref{Fig:VertexCorrection}(d) results in the correction
\begin{eqnarray}
\delta \lambda_{3}^{(4)} &=& -2N\frac{\lambda_{3}^{2}}{N}\int'
\frac{d\omega}{2\pi}\frac{d^{3}\mathbf{k}}{(2\pi)^{3}}
\mathrm{Tr}\left[\sigma_{3}G_{0}\left(i\omega,\mathbf{k}\right)\sigma_{3}\right.\nonumber
\\
&&\left.\times G_{0}\left(i\omega,\mathbf{k}\right)\right]\nonumber
\\
&=& \lambda_{3}^{2}
\frac{\Gamma\left(\frac{5}{4}\right)\Gamma\left(\frac{4}{3}\right)
\Lambda^{1/6}}{3\pi^{2}\Gamma\left(\frac{19}{12}\right)
\sqrt{A}B^{2/3}}\ell.
\end{eqnarray}
The correction induced by Fig.~\ref{Fig:VertexCorrection}(e) reads as
\begin{eqnarray}
\delta \lambda_{3}^{(5)} \sigma_{3} &=& -2g^{2}\lambda_{3}\int'
\frac{d\Omega}{2\pi}\frac{d^3\mathbf{q}}{(2\pi)^{3}}
G_{0}(i\Omega,\mathbf{q})\sigma_{3}G_{0}(i\Omega,\mathbf{q})\nonumber
\\
&&\times D_{0}(i\Omega,\mathbf{q}). \label{Eq:VertexCorrectionTQPTPara5}
\end{eqnarray}
Substituting Eqs.~(\ref{Eq:FermionPropagator}) and
(\ref{Eq:BosonPropagator}) into
Eq.~(\ref{Eq:VertexCorrectionTQPTPara5}), we arrive at
\begin{eqnarray}
\delta \lambda_{3}^{(5)} = \lambda_{3}C_{4}^{\prime}\ell,
\end{eqnarray}
where
\begin{eqnarray}
C_{4}^{\prime} &=& \frac{g^{2}}{12\pi^{2}\sqrt{A\Lambda}}
\int_{0}^{+\infty}d\xi\frac{\xi^{5/3}}{\left(1+
\xi^{2}\right)^{13/12}}\nonumber
\\
&&\times\frac{1}{\xi^{2/3}
\left(1+\xi^{2}\right)^{1/6} +\zeta}.
\end{eqnarray}
Diagrams shown in Figs.~\ref{Fig:VertexCorrection}(f) and \ref{Fig:VertexCorrection}(g) lead to
\begin{eqnarray}
V_{3}^{(6)+(7)} &=& -4g^{2}\lambda_{3}\int'\frac{d\Omega}{2\pi}
\frac{d^3\mathbf{q}}{(2\pi)^{3}}\left(\psi_{\sigma}^{\dag}
\sigma_{3} G_{0}(i\Omega,\mathbf{q})\psi_{\sigma}\right)\nonumber
\\
&&\times\left\{\psi^{\dag} \left[G_{0}(i\Omega,\mathbf{q})\sigma_{3} +
\sigma_{3} G_{0}(-i\Omega,-\mathbf{q})\right]\psi\right\}\nonumber
\\
&&\times D_{0}(i\Omega,\mathbf{q})\nonumber
\\
&=& -8\lambda_{3}g^{2}\int'\frac{d\Omega}{2\pi} \frac{d^3
\mathbf{q}}{(2\pi)^{3}}\frac{B^{2}d_{1}^{2}(\mathbf{k})}
{\left(\Omega^{2}+E_{\mathbf{k}}^{2}\right)^{2}}D_{0}(i\Omega,\mathbf{q})\nonumber
\\
&&\times
\left(\psi^{\dag}\sigma_{2}\psi\right)^{2}\nonumber
\\
&&-8\lambda_{3}g^{2}\int'\frac{d\Omega}{2\pi} \frac{d^3
\mathbf{q}}{(2\pi)^{3}}\frac{B^{2}d_{2}^{2}(\mathbf{k})}
{\left(\Omega^{2}+E_{\mathbf{k}}^{2}\right)^{2}}D_{0}(i\Omega,\mathbf{q})\nonumber
\\
&&\times \left(\psi^{\dag}
\sigma_{1}\psi\right)^{2}\nonumber
\\
&&-8\lambda_{3}g^{2}\int'\frac{d\Omega}{2\pi}
\frac{d^3\mathbf{q}}{(2\pi)^{3}} \frac{A^{2}
d_{3}^{2}(\mathbf{k})}{\left(\Omega^{2}+E_{\mathbf{k}}^{2}
\right)^{2}}D_{0}(i\Omega,\mathbf{q})\nonumber
\\
&&\times \left(\psi^{\dag}
\psi\right)^{2}.
\end{eqnarray}
Using Fierz identity and carrying out the integrations, we obtain
\begin{eqnarray}
V_{3}^{(6)+(7)} = \lambda_{3}C_{5}^{\prime}\ell \left(\psi^{\dag}
\sigma_{3}\psi\right)^{2},
\end{eqnarray}
where
\begin{eqnarray}
C_{5}^{\prime} &=& \frac{g^{2}}{6\pi^{2}\sqrt{A\Lambda}}
\int_{0}^{+\infty}d\xi \frac{1-\xi^{2}}{\xi^{1/3}
\left(1+\xi^{2}\right)^{13/12}}\nonumber
\\
&&\times \frac{1}{\xi^{2/3}
\left(1+\xi^{2}\right)^{1/6} +\zeta}.
\end{eqnarray}
This means that
\begin{eqnarray}
\delta\lambda_{3}^{(6)+(7)} = \lambda_{3}C_{5}^{\prime}\ell.
\end{eqnarray}
The correction for $\lambda_{3}$ from
Fig.~\ref{Fig:VertexCorrection}(h) is
\begin{eqnarray}
\delta\lambda_{3}^{(8)} &=& 2\lambda_{3}g^{2}\int'\frac{d\omega}{2\pi}
\frac{d^3\mathbf{k}}{(2\pi)^{3}} \mathrm{Tr}\left[G_{0}(i\omega,
\mathbf{k})\sigma_{3}\right.\nonumber
\\
&&\left.\times G_{0}(i\omega+i\Omega,\mathbf{k} +
\mathbf{q})\right]D_{0}(i\Omega,\mathbf{q})\nonumber
\\
&=&0.
\end{eqnarray}
Figs.~\ref{Fig:VertexCorrection}(i) and \ref{Fig:VertexCorrection}(j) yield
\begin{eqnarray}
V_{3}^{(9)+(10)} &=& 4g^{4}\int'\frac{d\Omega}{2\pi}
\frac{d^3\mathbf{q}}{(2\pi)^{3}} \left(\psi^{\dag}
G_{0}\left(i\Omega,\mathbf{q}\right) \psi\right)\nonumber
\\
&&\times D_{0}(i\Omega,\mathbf{q})\big\{\psi^{\dag}
\left[G_{0}\left(i\Omega,\mathbf{q}\right)\right.\nonumber
\\
&&\left.+G_{0}(-i\Omega,-\mathbf{q})\right]\psi\big\}
D_{0}(i\Omega,\mathbf{q}) \nonumber
\\
&=&\frac{\sqrt{A}B^{2/3}}{\Lambda^{1/6}}C_{6}
\ell\left(\psi^{\dag}\sigma_{3}
\psi\right)^{2},
\label{Eq:VertexCorrectionTQTP9and10}
\end{eqnarray}
where $C_{6}$ is expressed by Eq.~(\ref{Eq:C6Expression}). Thus,
\begin{eqnarray}
\delta\lambda_{3}^{(9)+(10)} = \frac{\sqrt{A}
B^{2/3}}{\Lambda^{1/6}}C_{6}\ell.
\end{eqnarray}

\begin{figure*}[htbp]
\center
\includegraphics[width=6.3in]{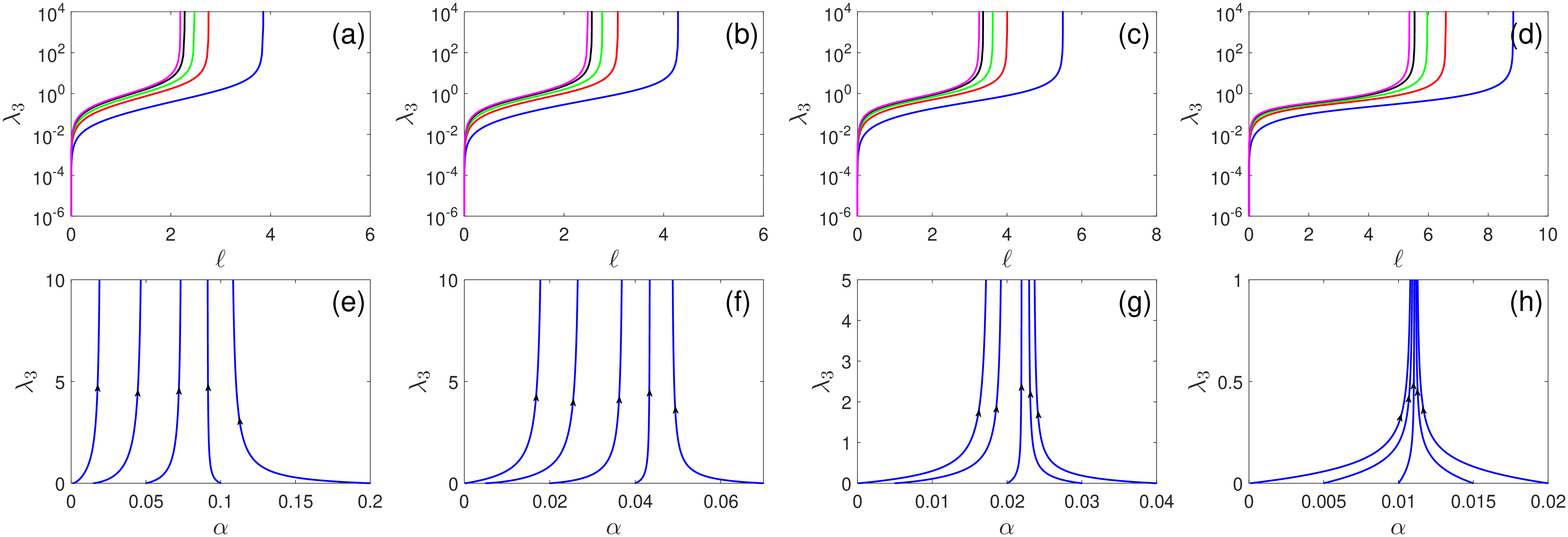}
\caption{(a)-(d): Dependence of $\lambda_{3}$ on $\ell$ obtained at
$N=1$, $N=2$, $N=4$, $N=8$, respectively. Blue, red, green, black,
magenta lines correspond to $\alpha_{0}=0.01, 0.05, 0.1, 0.2, 0.3$,
respectively. (e)-(h): Flow diagrams on the $\alpha$-$\lambda_{3}$
plane at $N=1$, $N=2$, $N=4$, and $N=8$, respectively. Here, we take
$\beta_{0}=0.1$. \label{Fig:FlowDiagramFFlambda3}}
\end{figure*}

The total correction can be written as
\begin{eqnarray}
\delta\lambda_{3} &=& \left(1-\frac{1}{4N}\right)\lambda_{3}^{2}
\frac{\Gamma\left(\frac{5}{4}\right) \Gamma\left(\frac{4}{3}\right)
\Lambda^{1/6}}{3\pi^{2}\Gamma\left(\frac{19}{12}\right)
\sqrt{A}B^{2/3}}\ell \nonumber
\\
&&+ \lambda_{3} \left(C_{4}^{\prime} +
C_{5}^{\prime}\right)\ell+\frac{\sqrt{A}
B^{2/3}}{\Lambda^{1/6}}C_{6}\ell.
\label{Eq:TotalCorrectionlambda3}
\end{eqnarray}
After incorporating the corrections, the action of four-fermion
interaction becomes
\begin{eqnarray}
S_{\psi^{4}}&=&\frac{1}{N}\left(\lambda_{3}+\delta\lambda_{3}\right)
\int\frac{d\omega_{1}}{2\pi}\frac{d^{3}\mathbf{k}_{1}}{(2\pi)^{3}}
\frac{d\omega_{2}}{2\pi}\frac{d^{3}\mathbf{k}_{2}}{(2\pi)^{3}}
\frac{d\omega_{3}}{2\pi}\frac{d^{3}\mathbf{k}_{3}}{(2\pi)^{3}}\nonumber
\\
&&\times \psi^{\dag}_{a}(\omega_{1},\mathbf{k}_{1})
\sigma_{3}\psi_{a}(\omega_{2},\mathbf{k}_{2})
\psi^{\dag}_{a}(\omega_{3},k_{3})\sigma_{3}\nonumber
\\
&&\times\psi_{a}(\omega_{1}-\omega_{2}+\omega_{3},
\mathbf{k}_{1}-\mathbf{k}_{2}+\mathbf{k}_{3}).
\end{eqnarray}
Utilizing the transformations
Eqs.~(\ref{Eq:ScalingOmega})-(\ref{Eq:Scalingpsi}), we arrive at
\begin{eqnarray}
S_{\psi'^{4}} &\approx& \frac{1}{N}\left(\lambda_{3}-\lambda_{3}
\frac{\ell}{6}+\delta\lambda_{3}\right)\int\frac{d\omega_{1}'}{2\pi}
\frac{d^{3}\mathbf{k}_{1}'}{(2\pi)^{3}}\frac{d\omega_{2}'}{2\pi}
\frac{d^{3}\mathbf{k}_{2}'}{(2\pi)^{3}}\nonumber
\\
&&\times \frac{d\omega_{3}'}{2\pi}
\frac{d^{3}\mathbf{k}_{3}'}{(2\pi)^{3}}
\psi'^{\dag}_{a}(\omega_{1}',\mathbf{k}_{1}')
\sigma_{3}\psi'_{a}(\omega_{2}',\mathbf{k}_{2}')
\psi'^{\dag}_{a}(\omega_{3}',k_{3}') \nonumber
\\
&&\times\sigma_{3}\psi'_{a}(\omega_{1}'-\omega_{2}'+\omega_{3}',
\mathbf{k}_{1}'-\mathbf{k}_{2}'+\mathbf{k}_{3}').
\end{eqnarray}
Defining a new parameter
\begin{eqnarray}
\lambda_{3}'=\lambda_{3}-\lambda_{3}\frac{\ell}{6} +
\delta\lambda_{3}, \label{Eq:Scalinglambda3}
\end{eqnarray}
we restore the original form of the action. From
Eqs.~(\ref{Eq:TotalCorrectionlambda3}) and
(\ref{Eq:Scalinglambda3}), one gets the flow equation for
$\lambda_{3}$
\begin{eqnarray}
\frac{d\lambda_{3}}{d\ell} &=& -\frac{1}{6}\lambda_{3} +
\frac{4}{5}\left(1-\frac{1}{4N}\right)\lambda_{3}^{2} +
\left(C_{4}^{\prime}+C_{5}^{\prime}-\frac{1}{2}C_{3}\right.\nonumber
\\
&&\left.-\frac{2}{3}C_{2}\right)\lambda_{3}+C_{7},
\end{eqnarray}
where
\begin{eqnarray}
C_{7} = \frac{5\Gamma\left(\frac{5}{4}\right)
\Gamma\left(\frac{4}{3}\right)}{12\pi^{2}\Gamma
\left(\frac{19}{12}\right)}C_{6}.
\end{eqnarray}
Once again, we have made the replacement
\begin{eqnarray}
\frac{5\Gamma\left(\frac{5}{4}\right)\Gamma\left(\frac{4}{3}\right)
\Lambda^{1/6}}{12\pi^{2}\Gamma\left(\frac{19}{12}\right)
\sqrt{A}B^{2/3}}\lambda_{3} \rightarrow \lambda_{3}.
\end{eqnarray}

\begin{figure*}[htbp]
\center
\includegraphics[width=6.3in]{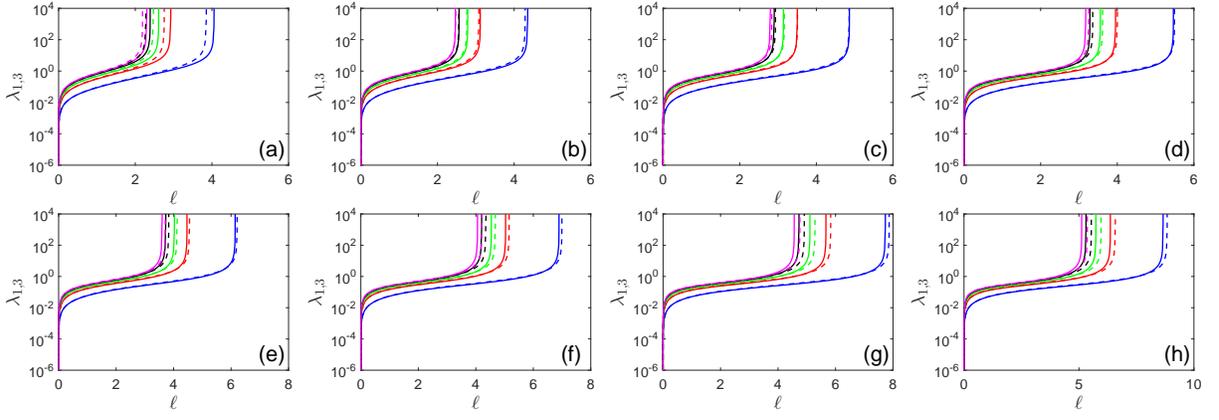}
\caption{Solid line corresponds to the flow of $\lambda_{1}$. Dashed
line represents the flow of $\lambda_{3}$. Blue, red, green, black,
and magenta lines correspond to $\alpha_{0}=0.01, 0.05, 0.1, 0.2$,
and $0.3$ respectively. In (a)-(h), the flavor is $N=1, 2, 3, 4, 5,
6, 7, 8$ respectively. Here, $\beta_{0}=0.1$.
\label{Fig:VRGLambda1Lambda3Compare}}
\end{figure*}

The RG flows of $\lambda_{3}$ for several possible values of fermion
flavor $N$ are presented in
Figs.~\ref{Fig:FlowDiagramFFlambda3}(a)-\ref{Fig:FlowDiagramFFlambda3}(d). The corresponding flow
diagrams on the $\alpha$-$\lambda_{3}$ plane are shown in
Figs.~\ref{Fig:FlowDiagramFFlambda3}(e)-\ref{Fig:FlowDiagramFFlambda3}(h). It is clear that
$\lambda_{3}$ grows from zero and becomes divergent at a finite
$\ell$ . Physically, these results indicate that
$\langle\psi^{\dag}\sigma_{3}\psi\rangle \neq 0$ due to the Coulomb
interaction. According to the above analysis, the TQPT between
triple-WSM and BI is continuous if
$\langle\psi^{\dag}\sigma_{3}\psi\rangle = 0$, but is first order
once $\langle\psi^{\dag}\sigma_{3}\psi\rangle$ becomes finite.

We now know that the system could either enter into a gapless
nematic phase or undergo a first order TQPT. It is necessary to
determine which instability is more favorable at low energies. For
this purpose, we now compare the RG flows of the coupling parameters
$\lambda_{1}$ and $\lambda_{3}$. According to the results shown in
Fig.~\ref{Fig:VRGLambda1Lambda3Compare}, we find that the Coulomb
interaction, within a wide range of initial values of $\alpha$,
drives $\lambda_{3}$ to diverge more quickly as the energy decreases
for fermion flavors $N=1$ and $N=2$. However, $\lambda_{1}$ becomes
divergent more quickly for flavors $N \geq 3$. We conclude that the
Coulomb interaction tends to trigger first order TQPT for $N=1,2$,
but leads to a gapless nematic state for $N \geq 3$.

\begin{figure*}[htbp]
\center
\includegraphics[width=6.3in]{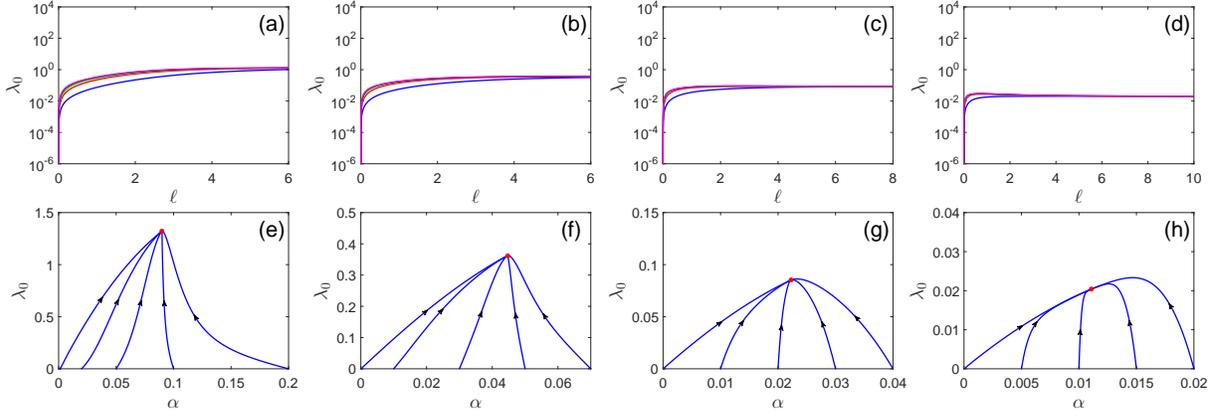}
\caption{(a)-(d): Dependence of $\lambda_{0}$ on $\ell$ with $N=1$,
$N=2$, $N=4$, $N=8$ respectively. Blue, red, green, black, and
magenta lines correspond to $\alpha_{0}=0.01, 0.05, 0.1, 0.2$, and
$0.3$ respectively. (e)-(h): Flow diagrams on the plane of the
parameters $\alpha$ and $\lambda_{0}$ with $N=1$, $N=2$, $N=4$, and
$N=8$ respectively. Here, we take $\beta_{0}=0.1$.
\label{Fig:FlowDiagramFFlambda0}}
\end{figure*}

\subsection{Interplay with $(\psi^{\dag}\psi)^{2}$}

The correction induced by Fig.~\ref{Fig:VertexCorrection}(a)
satisfies
\begin{eqnarray}
\delta \lambda_{0}^{(1)} &=& \frac{4\lambda_{0}^{2}}{N}\int'
\frac{d\omega}{2\pi}\frac{d^3\mathbf{k}}{(2\pi)^{3}}
G_{0}\left(i\omega,\mathbf{k}\right)
G_{0}\left(i\omega,\mathbf{k}\right)\nonumber
\\
&=&0.
\end{eqnarray}
The correction resulting from Figs.~\ref{Fig:VertexCorrection}(b)
and \ref{Fig:VertexCorrection}(c) takes the form
\begin{eqnarray}
V_{0}^{(2)+(3)} &=& \frac{4\lambda_{0}^{2}}{N}\int'\frac{d\omega}{2\pi}
\frac{d^3\mathbf{k}}{(2\pi)^{3}}\left(\psi^{\dag}
G_{0}(i\omega,\mathbf{k})\psi\right)\nonumber
\\
&&\times\left\{\psi^{\dag}
\left[G_{0}(i\omega,\mathbf{k})
+G_{0}(-i\omega,-\mathbf{k})\right]\psi\right\}\nonumber
\\
&=&\frac{\lambda_{0}^{2}}{N}\frac{\Gamma\left(\frac{1}{3}\right)
\Gamma\left(\frac{5}{4}\right)\Lambda^{1/6}}{12\pi^{2}
\Gamma\left(\frac{19}{12}\right)\sqrt{A}B^{2/3}}\ell
\left(\psi^{\dag}\sigma_{3}\psi\right)^{2},
\end{eqnarray}
which is equivalent to
\begin{eqnarray}
V_{0}^{(2)+(3)} = -\frac{\lambda_{0}^{2}}{N}
\frac{\Gamma\left(\frac{1}{3}\right)\Gamma\left(\frac{5}{4}\right)
\Lambda^{1/6}}{12\pi^{2}\Gamma\left(\frac{19}{12}\right)
\sqrt{A}B^{2/3}}\ell \left(\psi^{\dag}\psi\right)^{2},
\end{eqnarray}
according to Fierz identity. Thus $\delta\lambda_{0}^{(2)+(3)}$ can
be written as
\begin{eqnarray}
\delta \lambda_{0}^{(2)+(3)} = -\frac{\lambda_{0}^{2}}{N}
\frac{\Gamma\left(\frac{1}{3}\right)\Gamma\left(\frac{5}{4}\right)
\Lambda^{1/6}}{12\pi^{2}\Gamma\left(\frac{19}{12}\right)
\sqrt{A}B^{2/3}}\ell.
\end{eqnarray}
Diagram shown in Fig.~\ref{Fig:VertexCorrection}(d) generates the correction
\begin{eqnarray}
\delta \lambda_{0}^{(4)} &=& -2N\frac{\lambda_{0}^{2}}{N}\int'
\frac{d\omega}{2\pi}\frac{d^{3}\mathbf{k}}{(2\pi)^{3}}
\mathrm{Tr}\left[G_{0}\left(i\omega,\mathbf{k}\right)
G_{0}\left(i\omega,\mathbf{k}\right)\right]\nonumber
\\
&=& 0.
\end{eqnarray}
The correction due to Fig.~\ref{Fig:VertexCorrection}(e) reads
\begin{eqnarray}
\delta \lambda_{0}^{(5)} &=& -2g^{2}\lambda_{0}\int'
\frac{d\Omega}{2\pi}\frac{d^3\mathbf{q}}{(2\pi)^{3}}
G_{0}(i\Omega,\mathbf{q})G_{0}(i\Omega,\mathbf{q})\nonumber
\\
&&\times D_{0}(i\Omega,\mathbf{q})\nonumber
\\
&=& 0.
\end{eqnarray}
Figs.~\ref{Fig:VertexCorrection}(f) and \ref{Fig:VertexCorrection}(g) yield
\begin{eqnarray}
V_{0}^{(6)+(7)} &=& -4g^{2}\lambda_{0}\int'\frac{d\Omega}{2\pi}
\frac{d^3\mathbf{q}}{(2\pi)^{3}} \left(\psi_{\sigma}^{\dag}
G_{0}(i\Omega,\mathbf{q})\psi_{\sigma}\right)\nonumber
\\
&&\times\left\{\psi^{\dag}
\left[G_{0}(i\Omega,\mathbf{q})+G_{0}(-i\Omega,-\mathbf{q})\right]
\psi\right\}\nonumber
\\
&&\times D_{0}(i\Omega,\mathbf{q})\nonumber
\\
&=&-\lambda_{0}C_{5}^{\prime\prime}\ell\left(\psi^{\dag}\sigma_{3}
\psi\right)^{2}.
\end{eqnarray}
Using Fierz identity, we re-express $V^{(6)+(7)}$ as
\begin{eqnarray}
V_{0}^{(6)+(7)} = \lambda_{0}C_{5}^{\prime\prime}\ell\left(\psi^{\dag}
\psi\right)^{2},
\end{eqnarray}
where
\begin{eqnarray}
C_{5}^{\prime\prime}&=&\frac{g^{2}}{6\pi^{2}\sqrt{A\Lambda}}
\int_{0}^{+\infty}d\xi\frac{1}{\xi^{1/3}
\left(1+\xi^{2}\right)^{13/12}}\nonumber
\\
&&\times\frac{1}{\xi^{2/3}\left(1+\xi^{2}\right)^{1/6}
+\zeta}.
\end{eqnarray}
Thus, we obtain
\begin{eqnarray}
\delta\lambda_{0}^{(6)+(7)} = \lambda_{0}C_{5}^{\prime\prime}\ell.
\end{eqnarray}
Fig.~\ref{Fig:VertexCorrection}(h) results in
\begin{eqnarray}
\delta\lambda_{0}^{(8)}&=&2\lambda_{0}Ng^{2}\int'
\frac{d\omega}{2\pi}\frac{d^3\mathbf{k}}{(2\pi)^{3}}
\mathrm{Tr}\left[G_{0}(i\omega,\mathbf{k})\right.\nonumber
\\
&&\left.\times G_{0}(i\omega+i\Omega,\mathbf{k}+\mathbf{q})\right]
D_{0}(i\Omega,\mathbf{q})\nonumber \\
&\approx& -2\lambda_{0}N\left(C_{\bot} +
\frac{C_{z}}{\eta}\right)\ell.
\end{eqnarray}
The corrections induced by Figs.~\ref{Fig:VertexCorrection}(i)
and \ref{Fig:VertexCorrection}(j) are
\begin{eqnarray}
V_{0}^{(9)+(10)} &=& 4g^{4}\int'\frac{d\Omega}{2\pi}
\frac{d^3\mathbf{q}}{(2\pi)^{3}}\left(\psi^{\dag}
G_{0}\left(i\Omega,\mathbf{q}\right)
\psi\right)\nonumber
\\
&&\times D_{0}(i\Omega,\mathbf{q})\left\{\psi^{\dag}
\left[G_{0}\left(i\Omega,\mathbf{q}\right)\right.\right.\nonumber
\\
&&\left.\left.+G_{0}(-i\Omega,-\mathbf{q})\right]\psi\right\}
D_{0}(i\Omega,\mathbf{q})\nonumber
\\
&\approx&\frac{\sqrt{A}B^{2/3}}{\Lambda^{1/6}}
C_{6}\ell \left(\psi^{\dag}\sigma_{3}
\psi\right)^{2},
\end{eqnarray}
which can be further written as
\begin{eqnarray}
V_{0}^{(9)+(10)} &=& -\frac{\sqrt{A}B^{2/3}}{\Lambda^{
1/6}}C_{6}\ell \left(\psi^{\dag} \psi\right)^{2}.
\end{eqnarray}
Accordingly, we find
\begin{eqnarray}
\delta\lambda_{0}^{(9)+(10)} = -\frac{\sqrt{A}
B^{2/3}}{\Lambda^{1/6}}C_{6}\ell,
\end{eqnarray}
where $C_{6}$ is given by Eq.~(\ref{Eq:C6Expression}).

The total correction to $\lambda_{0}$ is
\begin{eqnarray}
\delta\lambda_{0} &=& -\frac{\lambda_{0}^{2}}{N} \frac{\Gamma
\left(\frac{1}{3}\right)\Gamma\left(\frac{5}{4}\right)
\Lambda^{1/6}}{12\pi^{2}\Gamma\left(\frac{19}{12}\right)
\sqrt{A}B^{2/3}}\ell + \lambda_{0}\bigg(C_{5}^{\prime
\prime}-2NC_{\bot}\nonumber
\\
&& - 2N\frac{C_{z}}{\eta}\bigg) \ell-\frac{\sqrt{A}
B^{2/3}}{\Lambda^{1/6}}C_{6}\ell.
\end{eqnarray}
We then obtain the following RG equation for $\lambda_{0}$
\begin{eqnarray}
\frac{d\lambda_{0}}{d\ell} &=& -\frac{1}{6}\lambda_{0} -
\frac{3}{5N}\lambda_{0}^{2} +\left(C_{5}^{\prime\prime}-2NC_{\bot} -
2N\beta\right) \lambda_{0}\nonumber
\\
&&- C_{7},
\end{eqnarray}
where
\begin{eqnarray}
C_{7}=\frac{5\Gamma\left(\frac{5}{4}\right)\Gamma\left(\frac{4}{3}\right)}
{12\pi^{2}\Gamma\left(\frac{19}{12}\right)}C_{6}.
\end{eqnarray}
The following re-definition
\begin{eqnarray}
\frac{5\Gamma\left(\frac{4}{3}\right)\Gamma\left(\frac{5}{4}\right)
\Lambda^{1/6}}{12\pi^{2}\Gamma\left(\frac{19}{12}\right)
\sqrt{A}B^{2/3}}\lambda_{0}\rightarrow\lambda_{0}
\end{eqnarray}
has been introduced during the derivation of RG equation.

The $\ell$-dependence of $\lambda_{0}$ can be seen from
Figs.~\ref{Fig:FlowDiagramFFlambda0}(a)-\ref{Fig:FlowDiagramFFlambda0}(d). Different from
$\lambda_1$ and $\lambda_3$, $\lambda_0$ flows to certain constant
as $\ell \rightarrow \infty$. The flow diagrams on the
$\alpha$-$\lambda_{0}$ are depicted in
Figs.~\ref{Fig:FlowDiagramFFlambda0}(e)-\ref{Fig:FlowDiagramFFlambda0}(h). Apparently, in this
case the interplay between Coulomb and the four-fermion interaction
does not lead to any instability of the system. The expectation
value $\langle\psi^{\dag}\psi\rangle$ always vanishes.

\end{document}